\documentclass[10pt,journal]{IEEEtran}
\IEEEoverridecommandlockouts
\usepackage{cite}
\usepackage{amsmath,amssymb,amsfonts}
\usepackage{algorithmic}
\usepackage{graphicx}
\usepackage{textcomp}
\usepackage{xcolor}
\usepackage[normalem]{ulem}
\usepackage{mathrsfs}
\usepackage{bm}
\usepackage[ruled,linesnumbered]{algorithm2e}
\usepackage{float}
\usepackage{url}
\usepackage{balance}
\usepackage{subfigure}
\usepackage{todonotes}
\usepackage{multirow}
\usepackage{tikz}
\usepackage{tabularx}
\usepackage{booktabs}
\usepackage{enumitem}
\newcommand{\revdel}[1]{}

\def\BibTeX{{\rm B\kern-.05em{\sc i\kern-.025em b}\kern-.08em
    T\kern-.1667em\lower.7ex\hbox{E}\kern-.125emX}}
\makeatletter
\renewcommand{\maketitle}{\par%
  \begingroup%
  \normalfont%
  \def\thefootnote{}%
  \def\footnotemark{}%
  \let\@makefnmark\relax%
  \footnotesize%
  \footnotesep 0.7\baselineskip%
  \normalsize%
  \twocolumn[{\IEEEquantizevspace{\@maketitle}[\IEEEquantizedisabletitlecmds]{0pt}[-\topskip]{\baselineskip}{\@IEEENORMtitlevspace}{\@IEEEMINtitlevspace}\@IEEEaftertitletext}]%
  \thispagestyle{IEEEtitlepagestyle}\@thanks%
  \endgroup%
  \setcounter{footnote}{0}\let\maketitle\relax\let\@maketitle\relax%
  \gdef\@thanks{}%
  \let\thanks\relax}
\makeatother
\begin{document}
\bstctlcite{BSTcontrol}
\title{AirTF: Over-the-Air Token Fusion for Task-Oriented Multi-Modal Token Communications}

\author{Bole Liu, Li Qiao, Minghui Wu, Yulin Shao, and Zhen Gao%
\vspace{-8mm}
\thanks{B. Liu, M. Wu, and Z. Gao are with the School of Information and Electronics, Beijing Institute of Technology, Beijing 100081, China (e-mail: \{leo2102, wuminghui, gaozhen16\}@bit.edu.cn). L. Qiao and Y. Shao are with the Department of Electrical and Computer Engineering, The University of Hong Kong, Pokfulam Road, Hong Kong (e-mail: \{qiaoli, ylshao\}@hku.hk).}}

\maketitle

\begin{abstract}

In the Internet of Vehicles (IoV), transmitting high-dimensional multi-modal sensory data to edge servers for time-sensitive tasks faces severe spectrum bottlenecks. To address this, we propose a foundation model-driven over-the-air token fusion (AirTF) framework for task-oriented multi-modal token communications. Unlike existing schemes for segmentation that rely on convolutional neural networks (CNNs) with limited local receptive fields, AirTF leverages vision transformer (ViT) encoders to extract globally contextualized semantic tokens from distributed heterogeneous sensors. By concurrently transmitting these spatially aligned tokens over a shared wireless channel, our framework exploits the superposition property of the multiple access channel to inherently fuse complementary multi-modal semantics (e.g., RGB and infrared) directly over the air. This mechanism significantly enhances spectral efficiency compared to orthogonal transmission. Furthermore, the integration of a pre-trained foundation model provides critical visual priors, effectively addressing the data-hungry nature of ViTs on limited, scenario-specific semantic segmentation datasets. Experiments demonstrate that AirTF consistently outperforms orthogonal transmission and CNN-based fusion baselines across AWGN and fading channels. Additional evaluations under a three-user setting, residual synchronization errors, and imperfect channel state information estimation further confirm its robustness.

\end{abstract}


\section{Introduction}
As the Internet of Vehicles (IoV) evolves toward autonomous driving, connected vehicles are increasingly equipped with heterogeneous sensors (e.g., cameras, infrared (IR) sensors, and LiDAR). To overcome the limited sensing range and computational power of individual vehicles, transmitting these multi-modal data to edge servers for collaborative inference becomes imperative~\cite{Roy2023multimodality,lu2025generativeAIIoV,Liu2023}. However, uploading such high-dimensional, high-volume raw sensory data imposes a severe bottleneck on limited vehicular spectrum resources. {\color{black}To address this challenge, semantic communication and over-the-air computation (AirComp) have emerged as two complementary paradigms for shifting communication systems from bit-level transmission toward semantic and task-oriented information delivery~\cite{wei2025taskorientedsemantic,qin2024aiemp}. On the one hand, DeepJSCC-based semantic communication learns compact, task-relevant representations and jointly maps them to wireless channels~\cite{bourtsoulatze2019deep,wu2024transformer,wu2024deepjsccmimo}. More recently, generative AI-enabled approaches further reduce communication overhead by conveying compact intents, pruning or merging redundant tokens, or extracting key features, while exploiting generative priors to reconstruct the omitted information~\cite{qiao2024latencyawaregenerative,liu2026communicatelesssynthesizerest,erak2025adaptive,wang2026kdsemnoma}. In parallel, token communication (TokenCom) is emerging as a large-model-native communication paradigm that treats model-native tokens, rather than conventional bit sequences, as the fundamental information-bearing and transmission units~\cite{qiao2025tokencom,liu2025texttoken,qiao2025todma}. Beyond compression, TokenCom exploits contextual priors and foundation-model reasoning for token-level inference and recovery, thereby mitigating the effects of packet loss, detection errors, and multiuser collisions while supporting cross-modal, text-guided, and multiuser communications. On the other hand, AirComp exploits the superposition property of wireless channels to compute nomographic functions, e.g., summation, of distributed data directly over the air, thereby alleviating uplink transmission pressure and saving communication bandwidth~\cite{zhu2021overtheair,Liu2025aircomp,qiao2026mdaircompplus}.

Combining these technologies is particularly critical for emerging IoV applications that pose strict demands on collaboration latency and efficiency~\cite{liu2024comfed}. Unlike the widely investigated image reconstruction tasks, autonomous driving requires task-oriented semantic communication for immediate decision-making, e.g., obstacle avoidance~\cite{zhu2021overtheair}, multi-modal behavior prediction~\cite{guo2025multimodal}, and occlusion-aware object detection~\cite{guo2024occlusion}.} In this context, recent studies~\cite{wu2025e2e,luo2024multimodal} have combined DeepJSCC with AirComp for task-oriented multi-user multi-modal sensing and inference. In such systems, modality-specific representations are transmitted concurrently and decoded at the edge server, allowing the received representation to combine color and texture cues from RGB images with thermal cues from IR images in low-light regions~\cite{sun2019rtfnet,shivakumar2020pst900}. Consequently, the optimization objective shifts from minimizing the mean squared error for signal reconstruction to maximizing the intersection over union for semantic segmentation. Under this task-oriented objective, end-to-end training encourages the received superposed representation to preserve complementary RGB and IR semantics for segmentation rather than reconstructing each modality at the pixel level. However, these works~\cite{wu2025e2e,luo2024multimodal} primarily employ convolutional neural networks (CNNs) for representation extraction and fusion. While CNNs are attractive for lightweight deployment, they still exhibit intrinsic limitations: Due to their local receptive fields, CNN-based methods may struggle to learn discriminative representations of semantic features~\cite{wu2024transformer}.

To address the above limitation and further improve communication efficiency, integrating vision foundation models, particularly vision transformers (ViTs)~\cite{dosovitskiy2020vit,liu2024toward,li2026largeModelEmbodied6g}, becomes essential. In ViTs, an image is represented as a sequence of patch tokens with positional embeddings. Specifically, we propose an over-the-air token fusion (AirTF) framework driven by such foundation models, where spatially aligned multi-modal tokens are directly superposed over the wireless channel to enable efficient semantic fusion. {\color{black}This architecture connects two complementary lines of research: token communication using Transformer architectures and pre-trained foundation models for semantic representation, and semantic AirComp for simultaneous multi-modal transmission and channel-level fusion.} The main contributions of this work are summarized as follows:

\begin{figure*}[ht]
    
    \centering
    \includegraphics[width=0.82\textwidth]{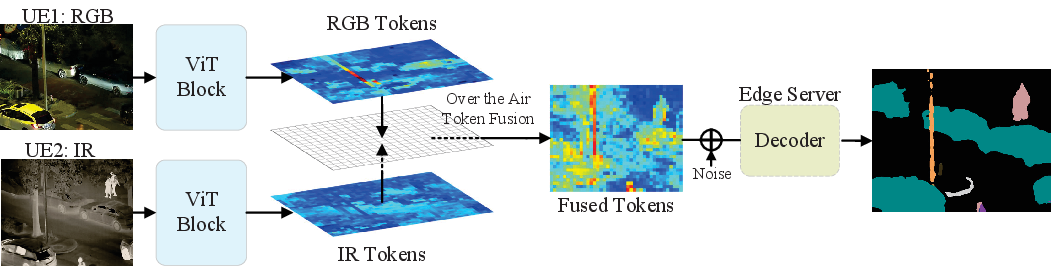}
    
    \caption{The overall architecture of the proposed multi-modal AirTF system. It consists of two modality-specific ViT encoders, over-the-air token fusion, and a unified semantic decoder. The token visualization depicts the magnitude of the generated tokens, where red indicates higher energy regions.}
    \label{fig:system_architecture}
    
\end{figure*}

\begin{itemize}
    \item {\color{black}We propose AirTF, an end-to-end multi-modal semantic communication framework that fuses the transmitted RGB and IR representations over a shared multiple access channel and directly predicts the semantic segmentation result at the edge server.}

    \item We develop a foundation model-driven transmission paradigm by fine-tuning {\color{black}ImageNet-initialized modality-specific ViT encoders}, which produce globally contextualized token representations and improve data efficiency compared to training the ViT backbone from scratch.

    \item Extensive experiments validate the effectiveness of the proposed framework.  At $\text{SNR} = 0$ dB and channel bandwidth ratio $C=1/256$ under AWGN channels, AirTF improves the mean intersection over union (mIoU) by 1.61\% and 2.07\% compared with the orthogonalized counterpart and the CNN-based baseline, respectively. Additional evaluations under Rayleigh and Rician fading and a three-user setting further demonstrate its effectiveness across different channel and modality configurations.

    \item We further evaluate AirTF under residual synchronization errors, imperfect channel state information (CSI) estimation, and spatial calibration errors, showing that it remains robust under moderate non-ideal conditions.
\end{itemize}

\textit{Notation}: Scalars, vectors, and matrices are denoted by italic lower-case, bold lower-case, and bold upper-case letters, respectively. The superscript $(\cdot)^T$ represents the transpose, $\mathbb{E}[\cdot]$ denotes the expectation, and $\|\cdot\|_F$ denotes the Frobenius norm.

\section{System Model}
As shown in Fig.~\ref{fig:system_architecture}, we consider an uplink task-oriented semantic communication scenario, where multiple user equipments (UEs) transmit their modality-specific sensory data to a central edge server (ES). Specifically, we focus on a two-user setting: UE 1 captures RGB images while UE 2 captures spatially aligned IR images of the same scene~\cite{luo2024multimodal}. In this case, we define $\mathcal{U}=\{\text{rgb},\text{ir}\}$ as the active UE set. The formulation can also be extended to a general case with $|\mathcal{U}|>2$ active modalities and UEs by assigning one modality-specific encoder and transmit stream to each UE. To maximize spectral efficiency in the primary setting, both UEs transmit their semantic features simultaneously over the same time-frequency resource blocks to the ES. 

In this work, a \textit{token} refers to a continuous-valued vector representation $\mathbf{z}_i \in \mathbb{R}^D$ output by the ViT encoder, where each token encodes the semantic information of a spatial image patch. The proposed AirTF framework leverages two modality-specific ViT encoders to extract semantic tokens from RGB and IR images locally. These tokens are then compressed and transmitted concurrently over the multiple access channel (MAC). As visualized in Fig.~\ref{fig:system_architecture}, the token heatmap reveals that end-to-end training drives the encoders to adaptively allocate higher transmission power to tokens corresponding to semantically critical regions (shown in red), such as pedestrians and vehicles, thereby ensuring their robustness against channel noise.

\subsection{Transmitter Architecture}
Let $\mathbf{X}_u \in \mathbb{R}^{3\times H\times W}$ denote the three-channel encoder input of UE $u \in \mathcal{U}$ after modality-specific preprocessing, where $(H,W)$ is the spatial resolution. The ViT backbone processes images as sequences of semantic tokens. For each modality branch $u\in\mathcal{U}$, the encoder input $\mathbf{X}_u$ is first reshaped into a sequence of flattened two-dimensional patches $\mathbf{x}_{p,u} \in \mathbb{R}^{N_p \times (3P^2)}$, where $P$ is the patch size and $N_p=\frac{HW}{P^2}$ is the number of patches.

These patches are mapped to a latent embedding space of dimension $D$ via a linear projection, yielding patch embeddings $\mathbf{E}_{p,u} \in \mathbb{R}^{N_p \times D}$. Following the standard ViT configuration, we define the input sequence length as $L = N_p + 1$. A learnable class token $\mathbf{z}_{\text{cls},u} \in \mathbb{R}^{D}$ is prepended to the sequence, and position embeddings $\mathbf{E}_{\text{pos},u} \in \mathbb{R}^{L \times D}$ are added to encode spatial structure
\begin{equation}
\mathbf{T}_u = [\mathbf{z}_{\text{cls},u}, \mathbf{E}_{p,u}] + \mathbf{E}_{\text{pos},u},
\end{equation}
where $[\cdot, \cdot]$ denotes concatenation along the token dimension. The resulting sequence $\mathbf{T}_u \in \mathbb{R}^{L \times D}$ is then fed into the transformer encoder
\begin{equation}
\mathbf{Z}_u = f_{\text{vit}}(\mathbf{T}_u; \boldsymbol{\theta}_u) \in \mathbb{R}^{L \times D},
\quad u\in\mathcal{U},
\end{equation}
where $f_{\text{vit}}(\cdot)$ represents the mapping function of the stacked transformer layers and $\boldsymbol{\theta}_u$ denotes the learnable parameters of the modality-specific ViT encoder for the $u$-th UE, $u\in\mathcal{U}$. Each output token vector $\mathbf{z}_i$ corresponds to a global context-aware representation centered around its original spatial location.

Considering the limited channel bandwidth, these high-dimensional tokens are projected into a compact transmission space and packed into complex channel symbols. Let 
$\mathcal{P}_{\rm IQ}: \mathbb{R}^{N_p\times d}\rightarrow 
\mathbb{C}^{N_p\times d/2}$ denote the I/Q packing operator for even $d$, where I/Q denotes the in-phase and quadrature components. 
The operator is defined as $[\mathcal{P}_{\rm IQ}(\mathbf{A})]_{n,m}=A_{n,2m-1}+jA_{n,2m}$ 
for any $\mathbf{A}\in\mathbb{R}^{N_p\times d}$, 
$n=1,\ldots,N_p$, and $m=1,\ldots,d/2$. The resulting complex token matrix is given by
\begin{equation}
\mathbf{S}_u
=
\mathcal{P}_{\rm IQ}
\left(
\left(\mathbf{Z}_u\mathbf{W}_u+\mathbf{b}_u\right)_{\mathrm{patch}}
\right)
\in \mathbb{C}^{N_p\times d/2},\quad u\in\mathcal{U},
\end{equation}

\noindent where $\mathbf{W}_u\in\mathbb{R}^{D\times d}$ and 
$\mathbf{b}_u\in\mathbb{R}^{d}$ are the learnable weight matrix and 
bias vector of the modality-specific dense projection layer, respectively, 
and $(\cdot)_{\rm patch}$ denotes retaining only the patch-token rows 
after removing the class token. With the CSI estimate $\widehat{h}_u$ 
available at the transmitters, each UE applies CSI-based phase 
pre-compensation and independent power normalization as

\begin{equation}
\tilde{\mathbf{S}}_u =
\sqrt{P_u N_s}
\frac{\mathbf{S}_u \mathrm{e}^{-j\angle \widehat{h}_u}}
{\|\mathbf{S}_u\|_F}.
\end{equation}

\noindent This operation uses the CSI estimate for phase pre-compensation, while the transmit power is normalized independently to satisfy the per-UE power constraint. The resulting transmit symbols satisfy
$\mathbb{E}[\|\tilde{\mathbf{S}}_u\|_F^2]/N_s \le P_u$ for each $u\in\mathcal{U}$, where $P_u$ denotes the average transmit-power budget per complex channel symbol for UE $u$. Accordingly, the number of shared complex channel uses per semantic frame is $N_s=N_p d/2$. In practice, the ES can provide a common timing and carrier reference through synchronization signals or pilots, and uplink pilots can be used to estimate residual channel phases~\cite{Shafi2017_5GTutorial, Garcia2021_NRV2X_Tutorial}. Each UE then applies timing advance and the phase pre-compensation in (4) before simultaneous transmission. {\color{black}AirTF assumes the availability of these network-assisted synchronization procedures. The effects of residual phase, timing, and CSI estimation errors after synchronization are evaluated in Section~\ref{subsec:robustness}.}

To quantify the communication overhead, we define the channel bandwidth ratio, denoted by $C$~\cite{bourtsoulatze2019deep, xu2022wireless}. It is calculated as the ratio of the number of shared complex channel uses $N_s$ to the dimension of the raw input source data, $C=N_s/(H \times W \times \sum_{u\in\mathcal{U}} C_u)$.
Here, $C_u$ denotes the number of raw source channels before modality-specific preprocessing. Hence, smaller $C$ implies a higher compression rate, indicating that fewer symbols are transmitted.

\subsection{Over-the-Air Token Fusion}
The multi-modal signals are transmitted over a shared MAC. At the ES,  the received complex baseband signal is modeled as
\begin{equation}
\mathbf{Y}
=
\sum_{u\in\mathcal{U}}
h_u\tilde{\mathbf{S}}_u
+
\mathbf{N},
\end{equation}
where $\mathbf{Y}\in\mathbb{C}^{N_p\times d/2}$ denotes the received complex-valued token matrix, $h_u$ denotes the complex channel coefficient of UE $u$, and $\mathbf{N}\in\mathbb{C}^{N_p\times d/2}$ has i.i.d. entries drawn from $\mathcal{CN}(0,\sigma_n^2)$, with $\sigma_n^2$ denoting the noise variance. 
We define the average transmit signal-to-noise ratio (SNR) under normalized fading as $\mathrm{SNR}=(\sum_{u\in\mathcal{U}} P_u)/\sigma_n^2$.
For fading channels, this SNR denotes the average transmit SNR, while the instantaneous received power can vary with fading realizations.

\subsection{Receiver Design}

At the receiver, the received complex token matrix 
$\mathbf{Y}\in\mathbb{C}^{N_p\times d/2}$ is unpacked as 
$\mathbf{Y}_{\rm r}=\mathcal{P}_{\rm IQ}^{-1}(\mathbf{Y})
\in\mathbb{R}^{N_p\times d}$, where inverse I/Q unpacking places the 
real and imaginary parts of each complex symbol into adjacent feature 
dimensions. The semantic decoder then maps $\mathbf{Y}_{\rm r}$ back 
to the spatial domain for segmentation. The decoding process consists of two stages: token refinement and spatial 
upsampling. In the first stage, this noise-corrupted sequence is projected 
back to the original embedding dimension $D$ and refined via a multi-layer 
perceptron (MLP) block with layer normalization and residual connections 
to mitigate the impact of channel noise. In the second stage, the refined tokens are reshaped into a 2D feature map $\mathbf{F}_{\text{2D}} \in \mathbb{R}^{D \times \frac{H}{P} \times \frac{W}{P}}$. The final segmentation map $\widehat{\mathbf{M}} \in \mathbb{R}^{K \times H \times W}$ is then generated through a decoder head comprising bilinear upsampling and convolutional layers
\begin{equation}
\widehat{\mathbf{M}} = g_{1\times1}\big( \phi ( \text{BN} ( g_{3\times3}( \operatorname{Up}(\mathbf{F}_{\text{2D}}) ) ) ) \big),
\end{equation}
where $K$ is the number of semantic classes, $\operatorname{Up}(\cdot)$ denotes bilinear upsampling, $g_{s \times s}(\cdot)$ represents a convolution operation with kernel size $s \times s$, $\text{BN}$ is batch normalization, and $\phi$ is the ReLU activation function.

\subsection{Pre-Training and Fine-Tuning}

Acquiring pixel-level annotations for semantic segmentation is labor-intensive, and ViTs are inherently data-hungry compared to CNNs due to the lack of inductive biases~\cite{dosovitskiy2020vit,cao2022training}, making training from scratch on small-scale semantic segmentation datasets challenging. To mitigate this challenge, we initialize the parameters of the RGB and IR ViT encoders with ImageNet-pre-trained weights~\cite{deng2009imagenet}. Although ImageNet is based on visible-light images, the learned generic visual priors remain useful for thermal imagery.

Subsequently, the entire system is fine-tuned in an end-to-end manner. The training objective is to minimize the discrepancy between the ground truth segmentation map $\mathbf{M}$ and the predicted map $\widehat{\mathbf{M}}$. For any spatial position $(i, j)$, the predicted logit vector is denoted by $\widehat{\mathbf{M}}_{:,i,j} \in \mathbb{R}^{K}$, and the corresponding one-hot ground truth label is given by $\mathbf{M}_{:,i,j} \in \{0,1\}^{K}$. We employ the cross-entropy loss function, defined as
\begin{equation}
\mathcal{L}_{\text{seg}} = -\sum_{i=1}^{H} \sum_{j=1}^{W} \sum_{k=1}^{K} \mathbf{M}_{k,i,j} \log \frac{\exp(\widehat{\mathbf{M}}_{k,i,j})}{\sum_{k'=1}^{K} \exp(\widehat{\mathbf{M}}_{k',i,j})},
\end{equation}
where $\mathbf{M}_{k,i,j}$ denotes the $k$-th element of the one-hot encoded label at position $(i,j)$.

{\color{black}We evaluate segmentation performance using an mIoU computed from the entries of the confusion matrix associated with the foreground classes. Let $\mathbf{C}$ denote the confusion matrix and $\mathcal{F}$ the set of foreground class indices, excluding the background class. The metric is defined as
\begin{equation}
\text{mIoU}=\frac{1}{|\mathcal{F}|}\sum_{k\in\mathcal{F}}
\frac{C_{kk}}
{\sum_{j\in\mathcal{F}}C_{kj}+\sum_{i\in\mathcal{F}}C_{ik}-C_{kk}},
\end{equation}
where $C_{ij}$ is the number of pixels whose ground-truth and predicted classes are $i$ and $j$, respectively. Both the class-wise union and the final average are computed over $\mathcal{F}$, and the same metric is applied uniformly to all compared schemes.}

\section{Experimental Results and Analysis}
\subsection{Experimental Setup}

\textbf{Dataset and Implementation:} We evaluate the proposed scheme on the SemanticRT dataset~\cite{ji2023semanticrt}, following the original data split. The dataset contains 11,371 aligned pairs of RGB and IR images with 13 semantic classes. The images have a resolution of $512 \times 640$ and cover complex automotive scenarios, including low-light and pitch-black environments. RGB and IR inputs are resized to $512\times640$; random scaling/cropping is used only for training augmentation, and single-channel IR images are converted to three-channel encoder inputs. For the multi-user evaluation, we use the PST900 dataset~\cite{shivakumar2020pst900}, which provides aligned RGB, thermal, and depth data.
The simulation is conducted on a workstation equipped with a single NVIDIA RTX 5090 GPU, using PyTorch 2.5.1. We adopt the \textbf{ViT-S/16} architecture~\cite{dosovitskiy2020vit} as the semantic backbone. For the fading-channel experiments, each UE is assigned one complex fading coefficient $h_u$ for each image sample, and this coefficient is kept fixed over all transmitted semantic symbols of that image. Rayleigh fading uses $h_u\sim\mathcal{CN}(0,1)$, while Rician fading follows the same normalized setup with a line-of-sight component controlled by the Rician $K$-factor. CSI is available at the transmitters, and each UE applies phase pre-compensation with per-UE power normalization.

We compare the proposed AirTF scheme with the following baselines: (1) \textbf{ViT-SemCom-OMA}: {\color{black}ViT-based DeepJSCC has been investigated for wireless image transmission~\cite{wu2024deepjsccmimo}. To provide a task-matched comparison under the considered multi-modal segmentation setting, we construct ViT-SemCom-OMA as an OMA baseline. This scheme employs the same modality-specific ViT encoder architecture and segmentation objective as AirTF. The RGB and IR semantic representations are transmitted using OMA. To maintain the same total channel bandwidth ratio, each modality is allocated half of the channel uses employed by AirTF. At the receiver, the recovered representations are concatenated and processed by the semantic decoder.} (2) \textbf{MFNet}: {\color{black}A representative CNN-based RGB--thermal fusion architecture~\cite{ha2017mfnet}, adapted to the same non-orthogonal transmission setting following the feature superposition paradigm in~\cite{luo2024multimodal}.}

\textbf{Training Strategy:}  The pre-trained AirTF and ViT-SemCom-OMA models in the main comparison are fine-tuned for 50 epochs, while the MFNet baseline is trained for 100 epochs. All models are optimized using AdamW with a weight decay of $1 \times 10^{-4}$ and a cosine annealing learning-rate scheduler. For AirTF and ViT-SemCom-OMA, ViT encoders initialized with ImageNet pre-trained weights use an initial learning rate of $1 \times 10^{-5}$, while the remaining trainable modules use $1 \times 10^{-4}$. The MFNet baseline uses an initial learning rate of $1 \times 10^{-4}$.

\begin{figure}[!t]
    
    \centering
    
    \includegraphics[width=0.92\columnwidth]{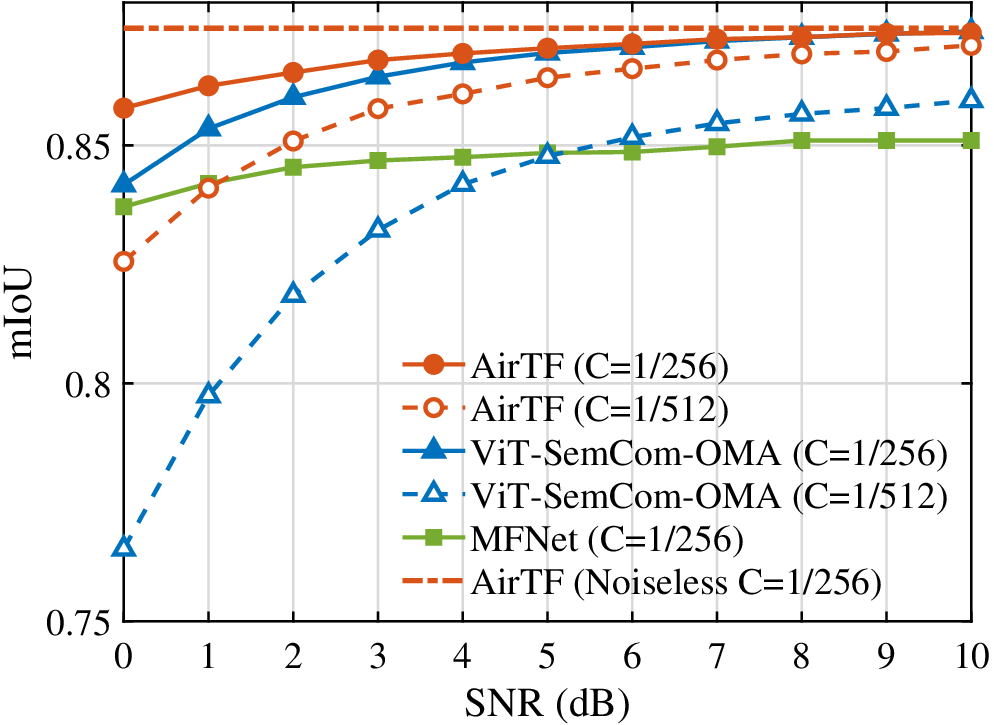}
    
    \caption{Semantic segmentation performance comparison versus SNR under AWGN channels.}
    \label{fig:performance_snr}
    
\end{figure}

\begin{figure*}[!t]
    
    \centering
    \includegraphics[width=0.90\textwidth]{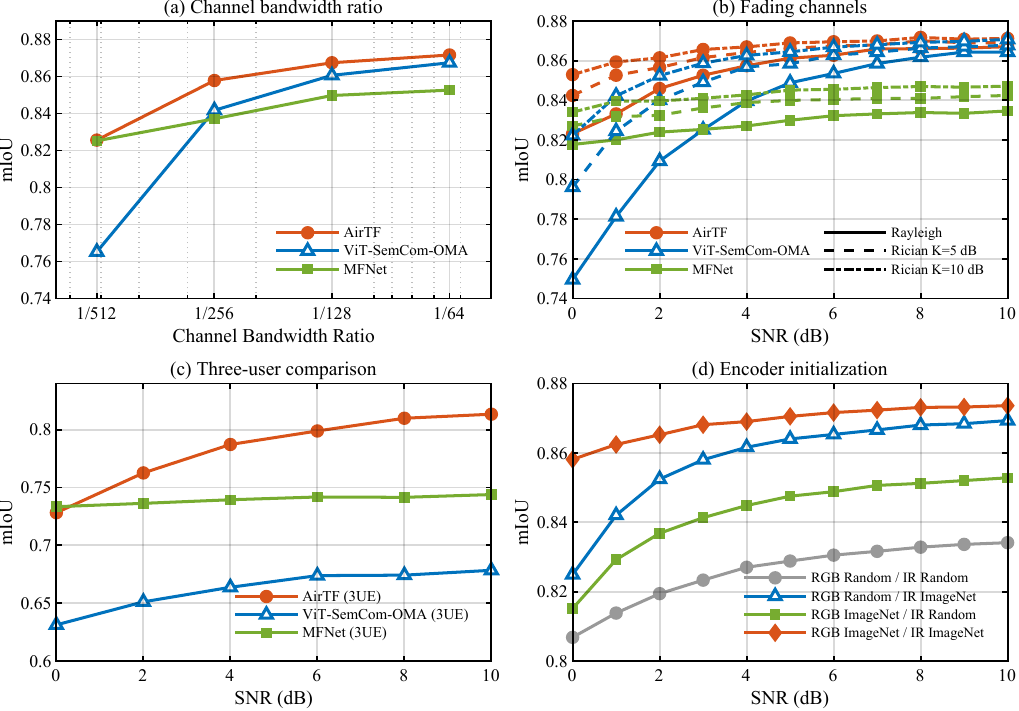}
    
    \caption{Quantitative results: (a) channel bandwidth ratio ablation under AWGN at SNR $=0$ dB; (b) performance under Rayleigh and Rician fading channels at $C=1/256$; (c) performance under AWGN on the PST900 three-user setting, where AirTF and MFNet use $C=1/256$ and ViT-SemCom-OMA uses $C\approx1/213.3$; and (d) encoder initialization ablation at $C=1/256$.}
    \label{fig:experimental_results}
    
\end{figure*}

\subsection{Performance Comparison}

We evaluate the system performance under varying AWGN channel conditions. As shown in Fig.~\ref{fig:performance_snr}, at SNR = 0 dB and $C=1/256$, AirTF achieves an mIoU of 85.78\%, which is 1.61\% higher than ViT-SemCom-OMA (84.17\%) and 2.07\% higher than MFNet (83.71\%). As SNR increases, ViT-SemCom-OMA gradually approaches AirTF at $C=1/256$. The noiseless AirTF curve is included as a reference performance ceiling of the same transmission architecture when channel noise is removed. For $C=1/512$, a clear performance gap persists, demonstrating the bandwidth efficiency of AirTF under limited channel resources. The performance gain of AirTF over MFNet is attributed to the advantage of the vision transformer architecture and the pre-training strategy. Specifically, while CNNs are limited by local receptive fields, ViT's self-attention mechanism captures global semantic context, enabling the generation of high-quality semantic tokens that retain critical scene information even after channel corruption.

Fig.~\ref{fig:experimental_results} further evaluates the effects of channel bandwidth ratio, fading channels, the PST900 three-user setting, and encoder initialization. Fig.~\ref{fig:experimental_results}(a) evaluates task performance against varying channel bandwidth ratios at SNR $=0$ dB. While all schemes exhibit a performance decline as the channel bandwidth ratio decreases, the orthogonal approach suffers a more pronounced degradation due to insufficient per-modality resources. In contrast, the proposed scheme maintains robust utility. At low channel bandwidth ratios, the performance gap between the Transformer-based AirTF and CNN-based MFNet becomes very small. This phenomenon suggests that uniform dimension reduction across all spatial tokens imposes a representation bottleneck, limiting the expression of semantically critical features under severe constraints. Consequently, integrating importance-aware token allocation into the proposed framework represents a promising avenue to further exploit the potential of ViTs. Nevertheless, the proposed over-the-air fusion mechanism effectively alleviates the severe degradation observed in orthogonal schemes, demonstrating superior robustness in preserving multi-modal semantics under bandwidth scarcity. Fig.~\ref{fig:experimental_results}(b) further evaluates the models under Rayleigh fading and Rician fading with $K$-factors of 5 and 10 dB. The curves show the expected trend that performance improves as the line-of-sight component becomes stronger, with Rayleigh fading serving as the no-line-of-sight limiting case. Across these fading settings, AirTF consistently achieves higher mIoU than ViT-SemCom-OMA, although the gap narrows at high SNR, and maintains a clear advantage over MFNet, indicating that AirTF remains effective beyond AWGN channels. Fig.~\ref{fig:experimental_results}(c) reports the PST900 multi-user experiment with RGB, thermal, and depth inputs, where AirTF and MFNet use $C=1/256$ and ViT-SemCom-OMA uses $C\approx1/213.3$. At SNR $=5$ dB, AirTF achieves 79.33\% mIoU, compared with 66.80\% for ViT-SemCom-OMA and 74.06\% for MFNet, showing that the proposed framework can extend to heterogeneous multi-user sensing scenarios. {\color{black}The two- and three-user experiments validate AirTF in small-scale multi-modal cooperative perception settings, while its extension to large-scale vehicular networks with user selection, resource scheduling, and mobility management will be investigated in future work.}

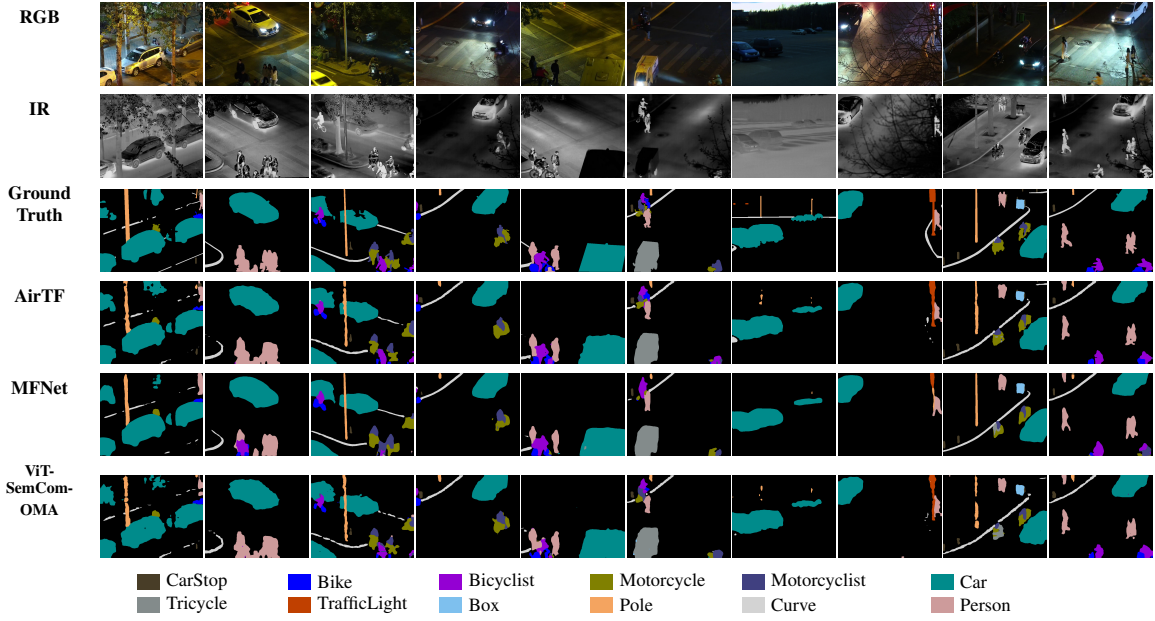
\begin{figure*}[ht]
    
    \centering
    \setlength{\tabcolsep}{0.5pt}

    \newcommand{\imgW}{0.075\textwidth}
    
    \newcommand{\rowL}[1]{\raisebox{2.3em}{\parbox{1.0cm}{\centering \scriptsize \bfseries #1}}}
    
    \newcommand{\figpath}{Fig4/fig4_images_cbr512_test/tiles}
    
    \begin{tabular}{c @{\hspace{3mm}} cccccccccc}
        
        \rowL{RGB} &
        \includegraphics[width=\imgW]{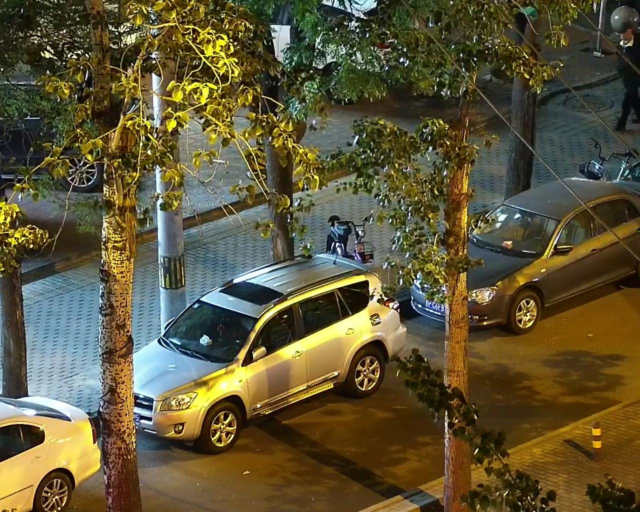} &
        \includegraphics[width=\imgW]{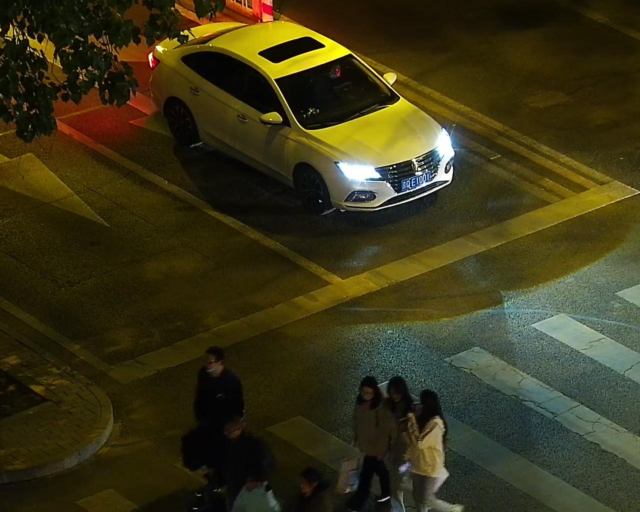} &
        \includegraphics[width=\imgW]{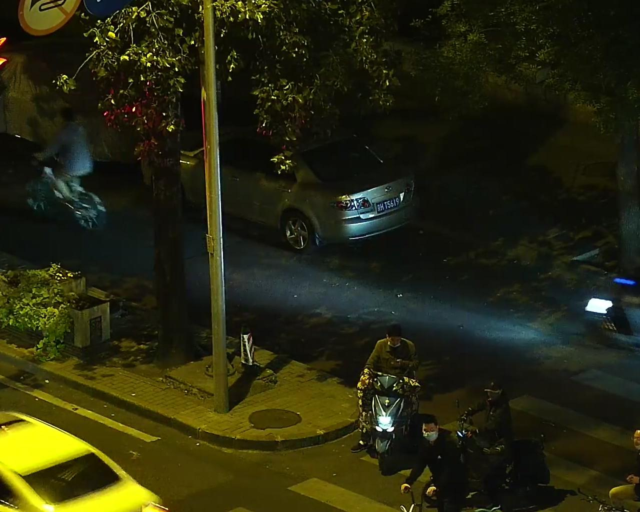} &
        \includegraphics[width=\imgW]{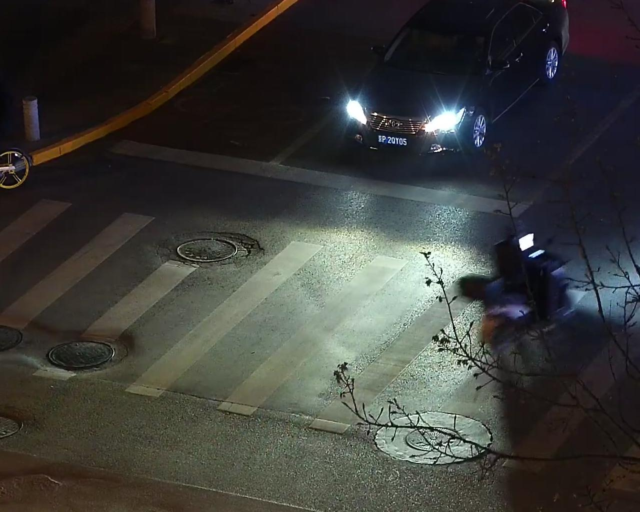} &
        \includegraphics[width=\imgW]{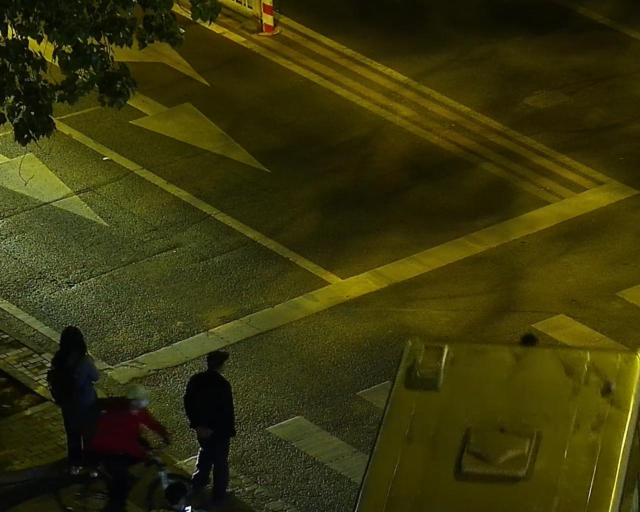} &
        \includegraphics[width=\imgW]{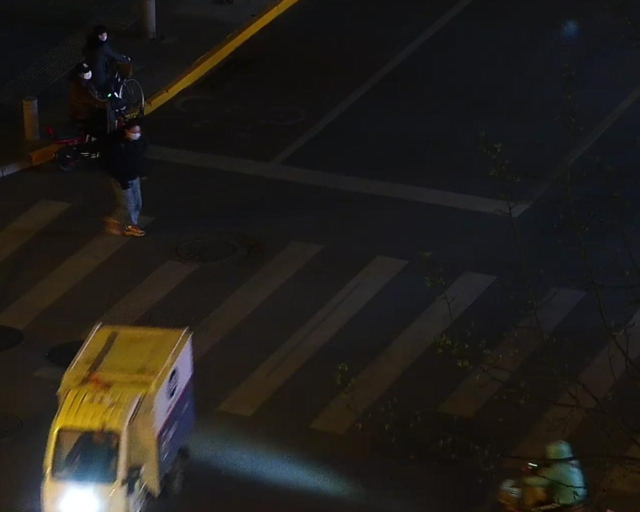} &
        \includegraphics[width=\imgW]{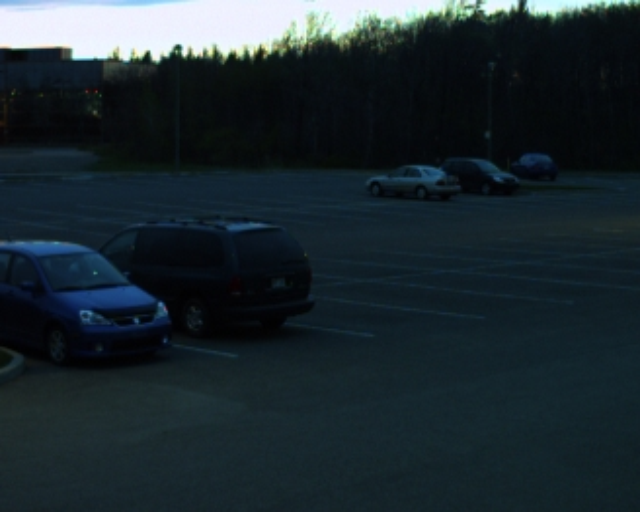} &
        \includegraphics[width=\imgW]{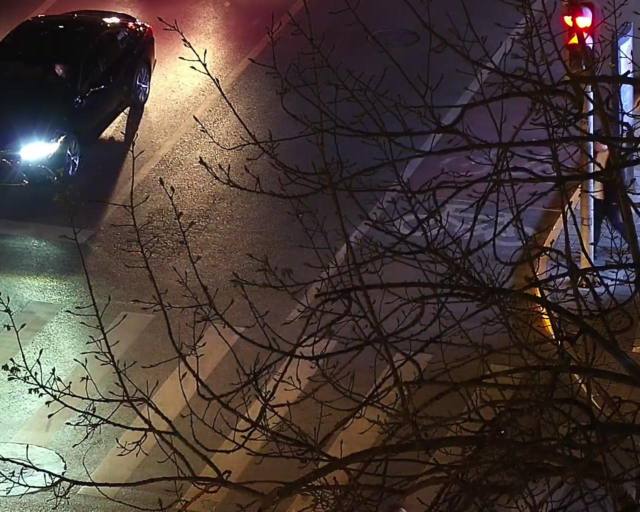} &
        \includegraphics[width=\imgW]{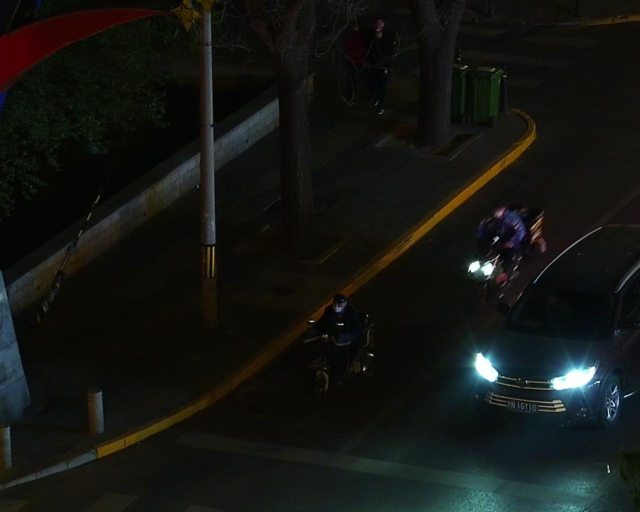} &
        \includegraphics[width=\imgW]{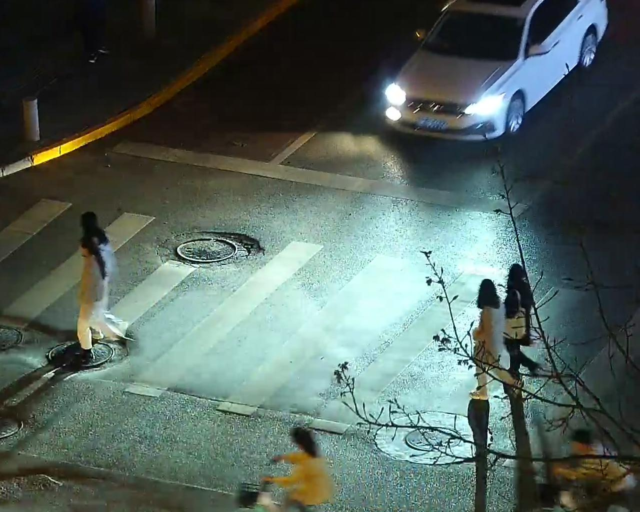} \\
        
        \rowL{IR} &
        \includegraphics[width=\imgW]{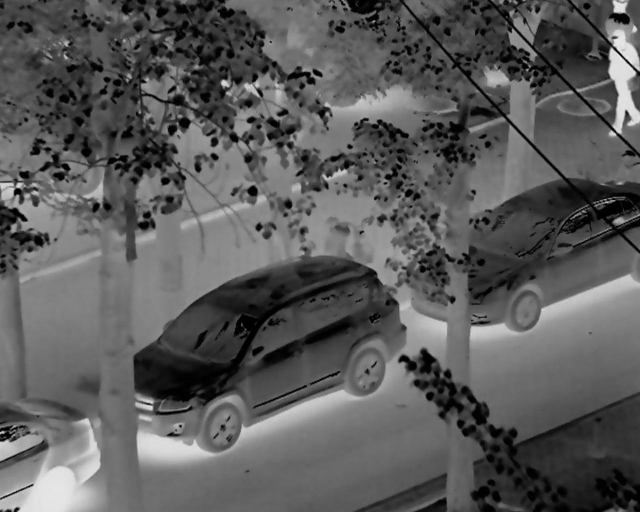} &
        \includegraphics[width=\imgW]{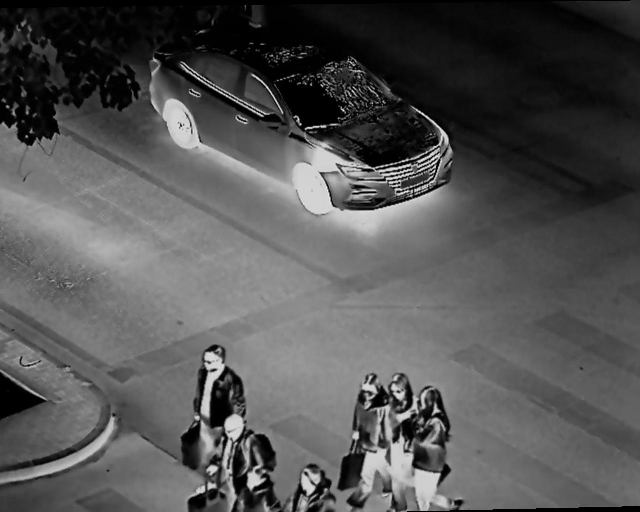} &
        \includegraphics[width=\imgW]{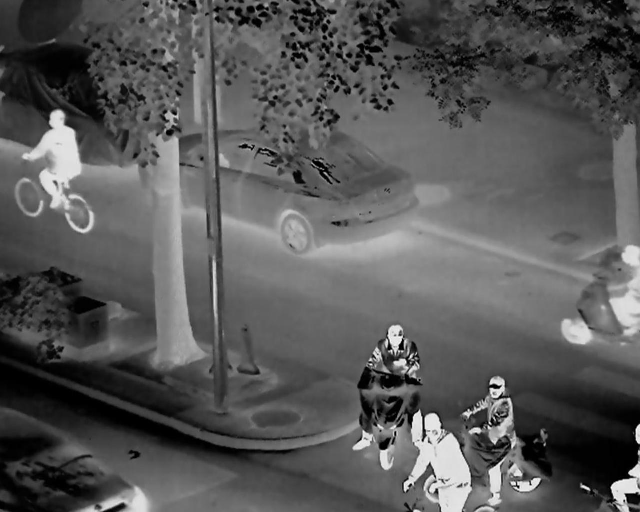} &
        \includegraphics[width=\imgW]{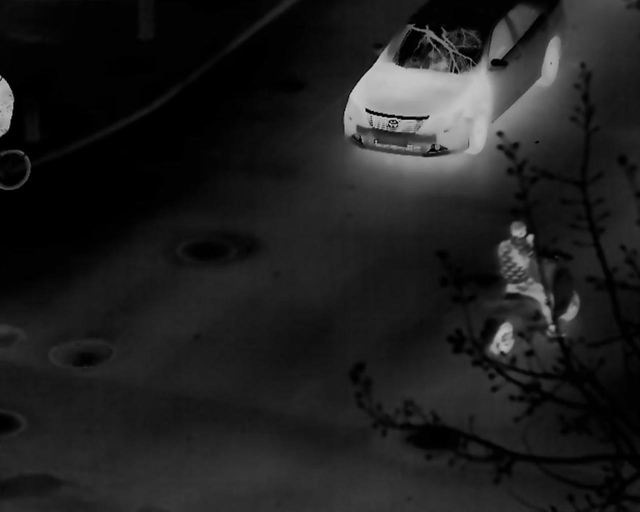} &
        \includegraphics[width=\imgW]{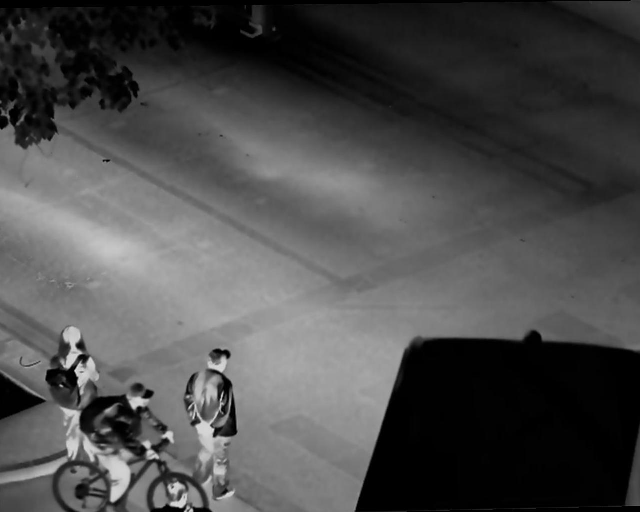} &
        \includegraphics[width=\imgW]{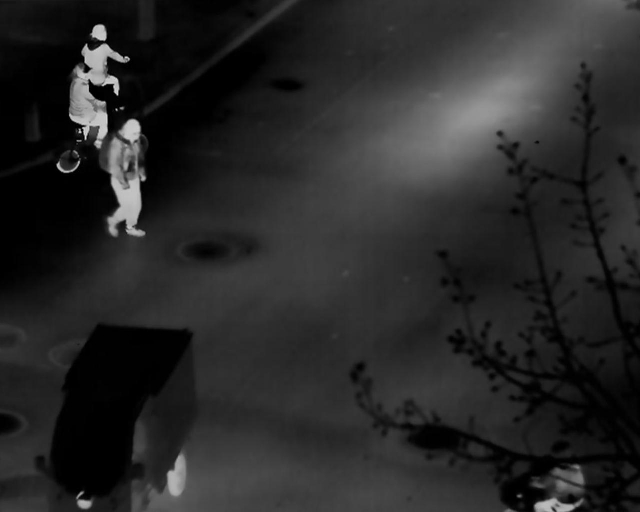} &
        \includegraphics[width=\imgW]{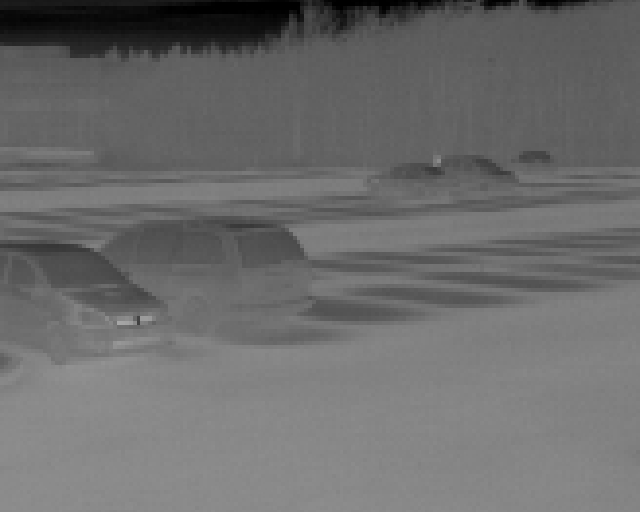} &
        \includegraphics[width=\imgW]{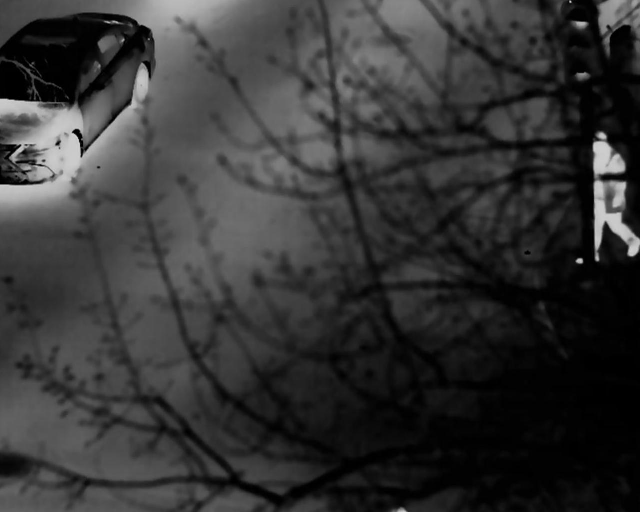} &
        \includegraphics[width=\imgW]{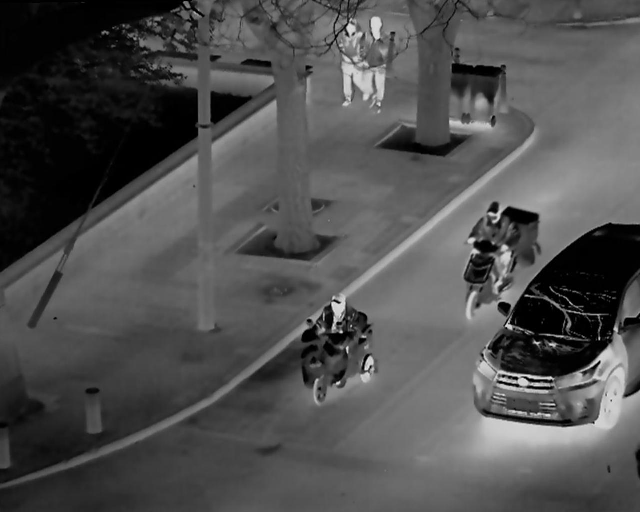} &
        \includegraphics[width=\imgW]{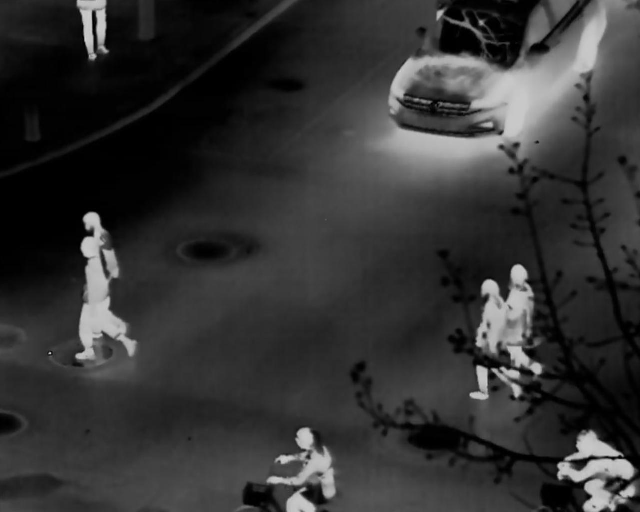} \\

        \rowL{Ground\\Truth} &
        \includegraphics[width=\imgW]{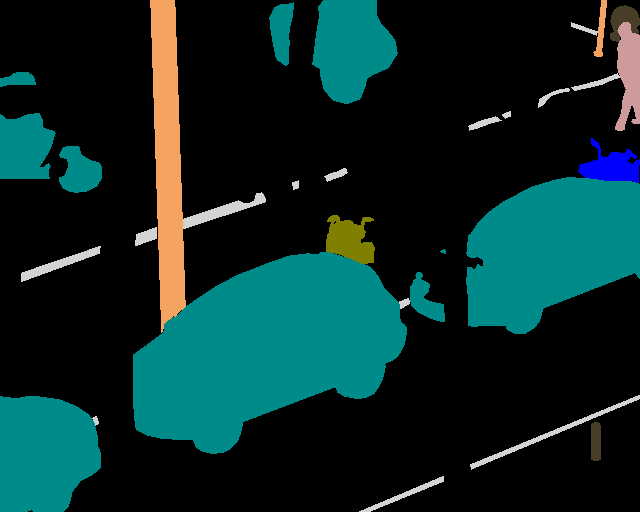} &
        \includegraphics[width=\imgW]{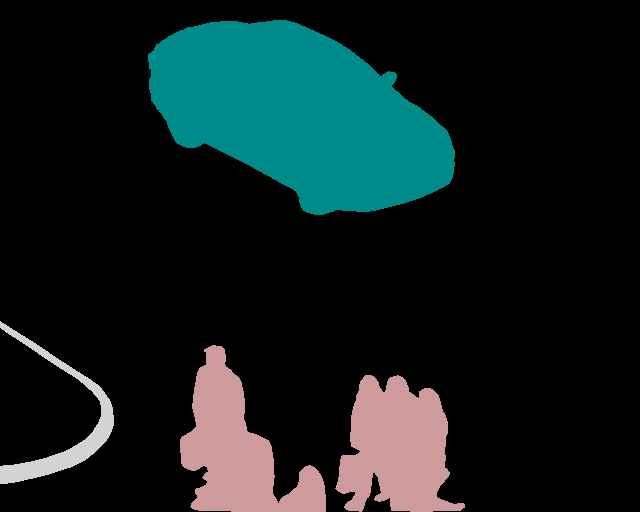} &
        \includegraphics[width=\imgW]{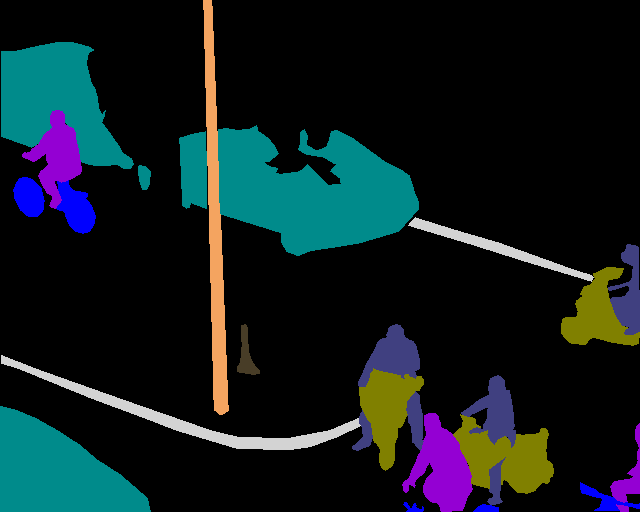} &
        \includegraphics[width=\imgW]{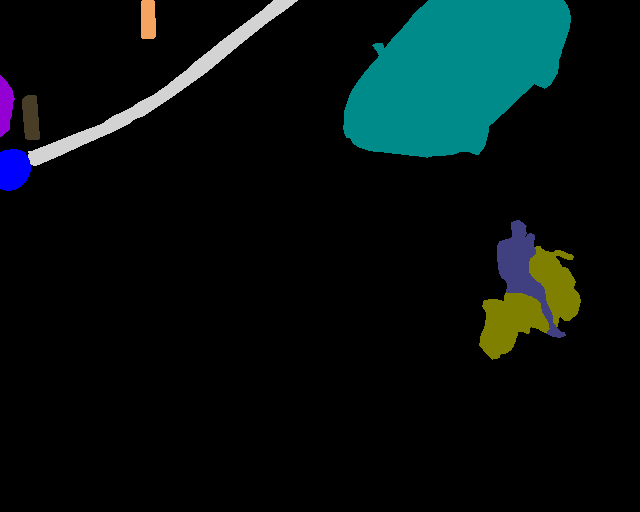} &
        \includegraphics[width=\imgW]{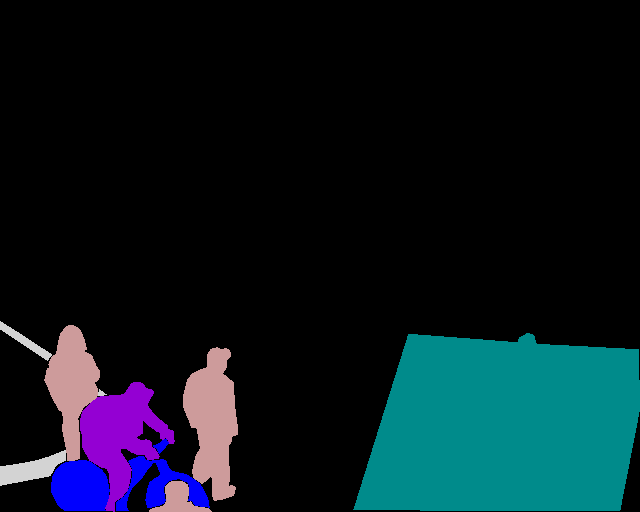} &
        \includegraphics[width=\imgW]{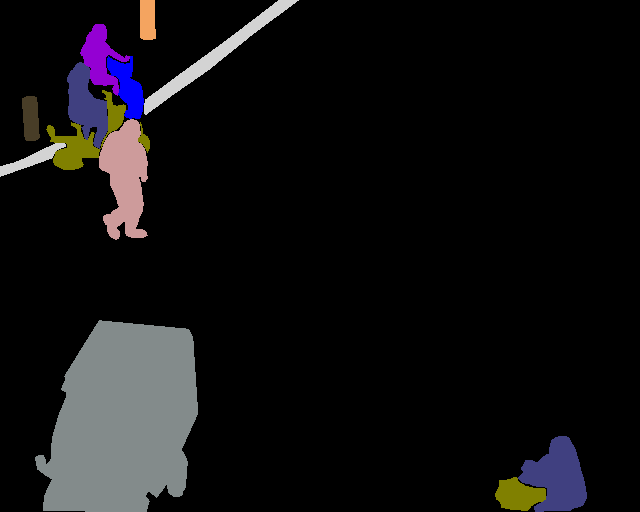} &
        \includegraphics[width=\imgW]{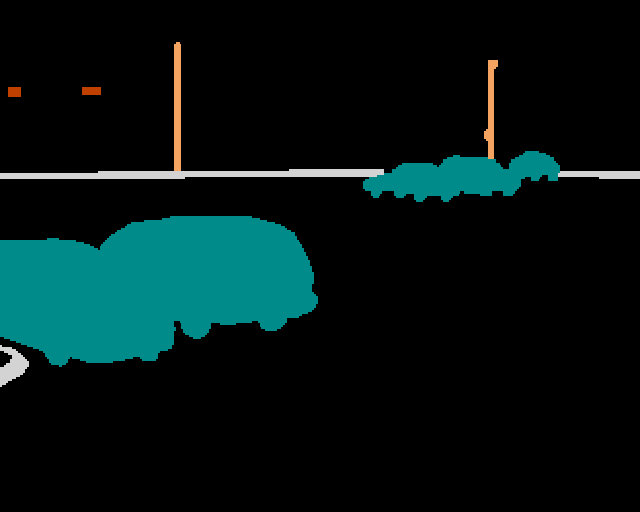} &
        \includegraphics[width=\imgW]{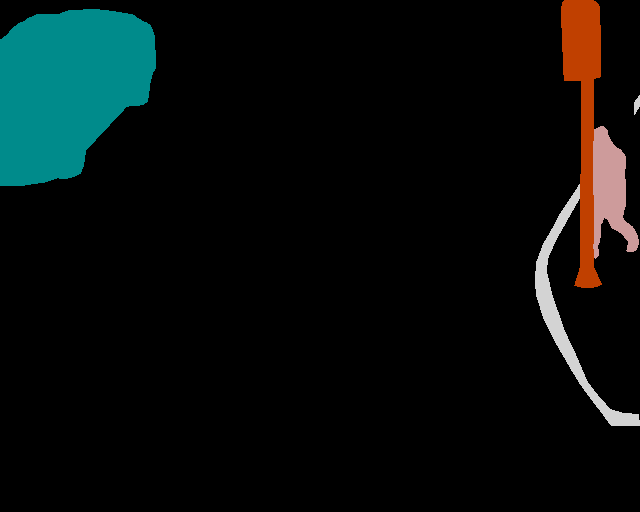} &
        \includegraphics[width=\imgW]{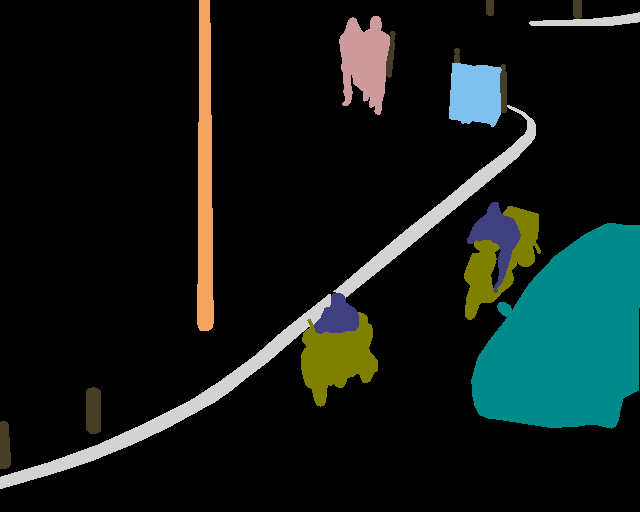} &
        \includegraphics[width=\imgW]{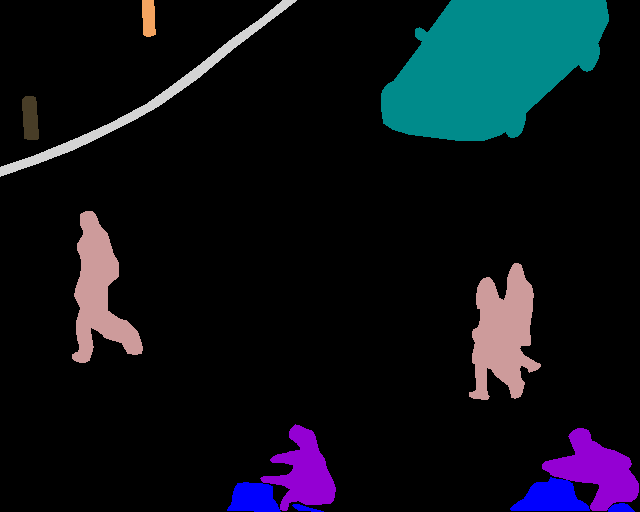} \\

        \rowL{AirTF} &
        \includegraphics[width=\imgW]{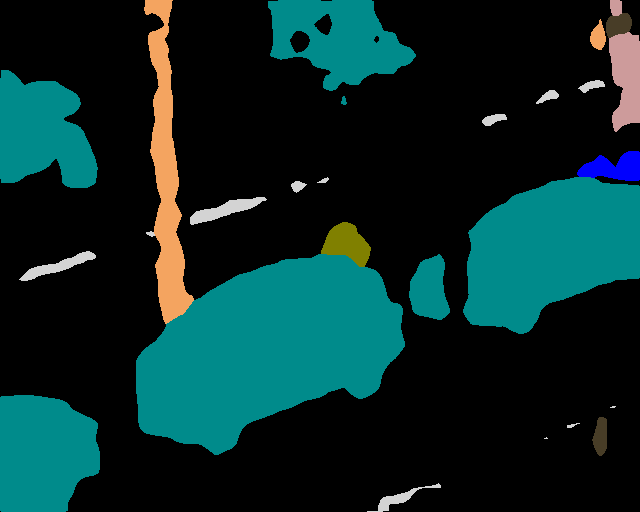} &
        \includegraphics[width=\imgW]{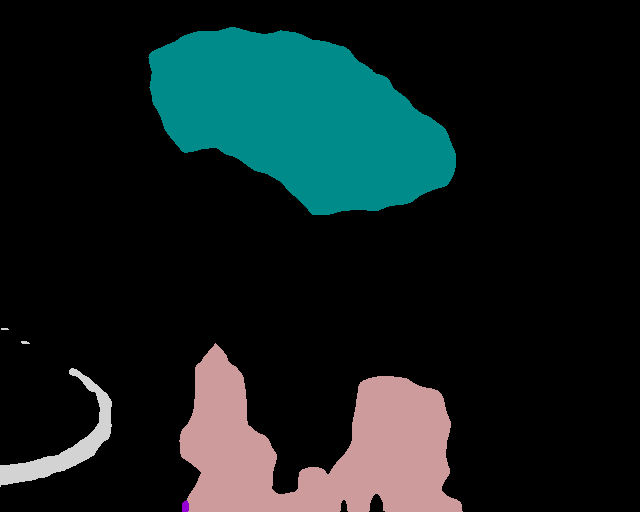} &
        \includegraphics[width=\imgW]{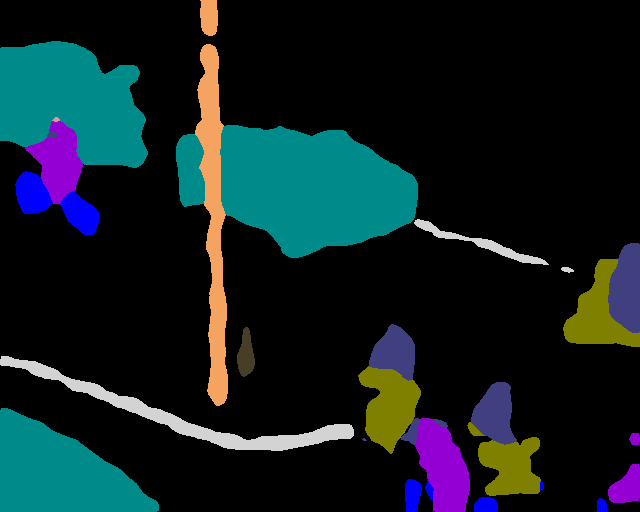} &
        \includegraphics[width=\imgW]{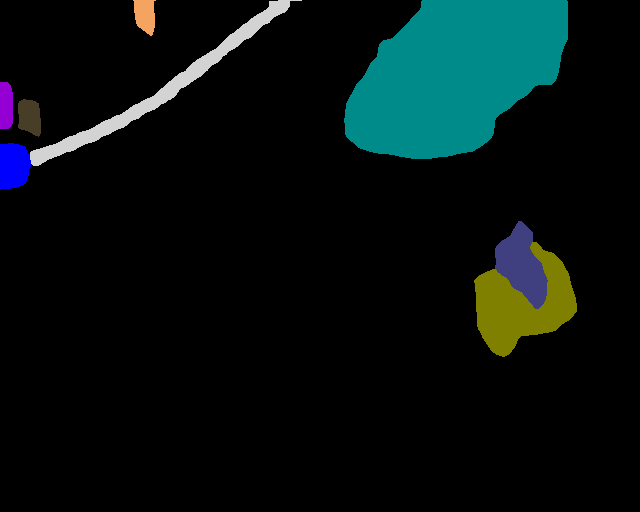} &
        \includegraphics[width=\imgW]{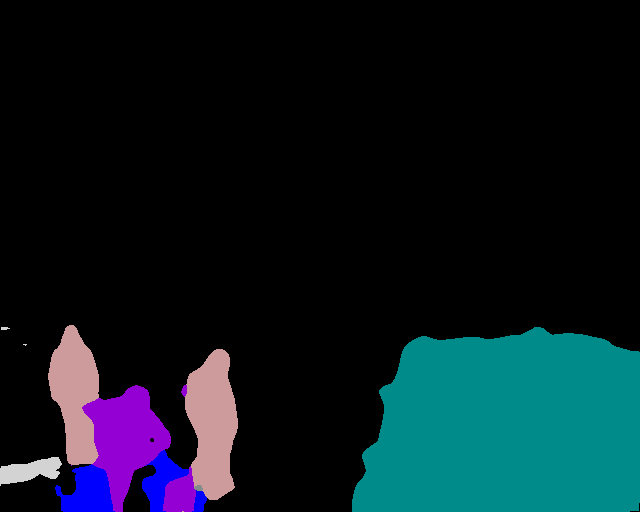} &
        \includegraphics[width=\imgW]{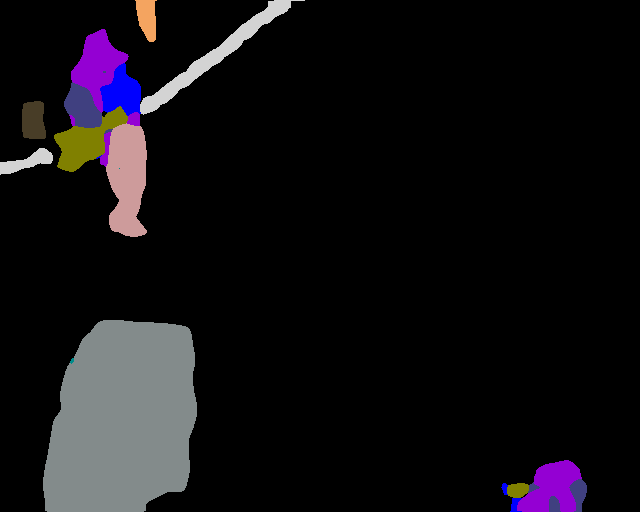} &
        \includegraphics[width=\imgW]{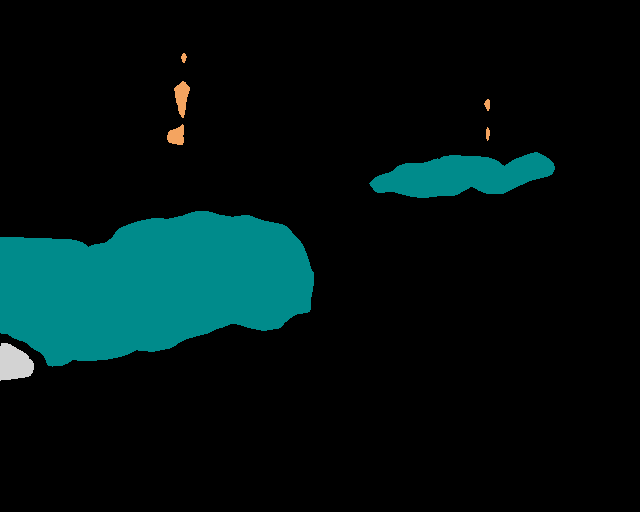} &
        \includegraphics[width=\imgW]{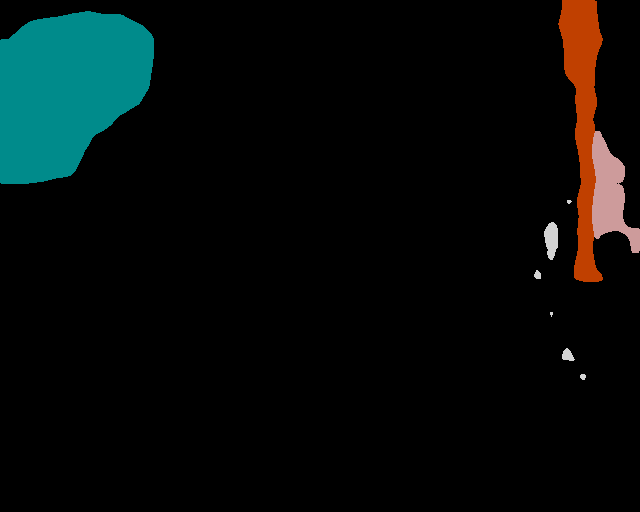} &
        \includegraphics[width=\imgW]{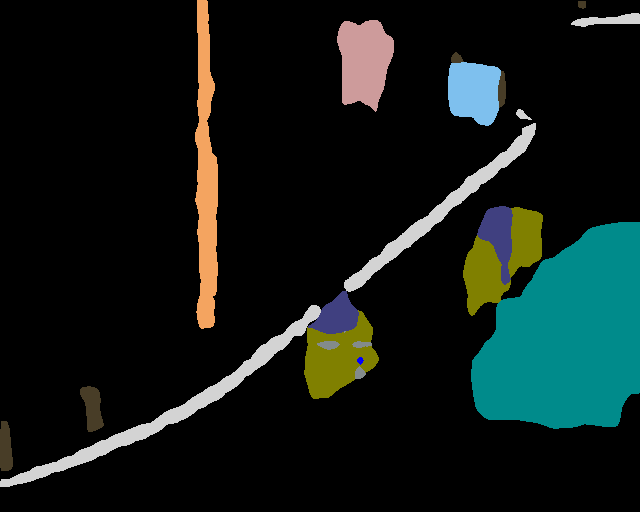} &
        \includegraphics[width=\imgW]{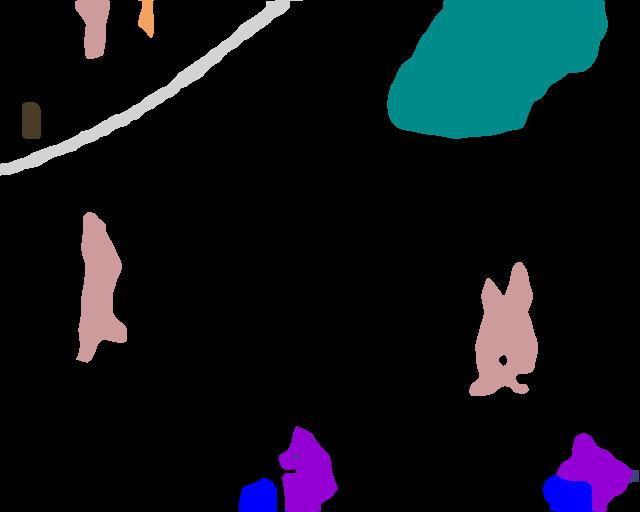} \\

        \rowL{MFNet} &
        \includegraphics[width=\imgW]{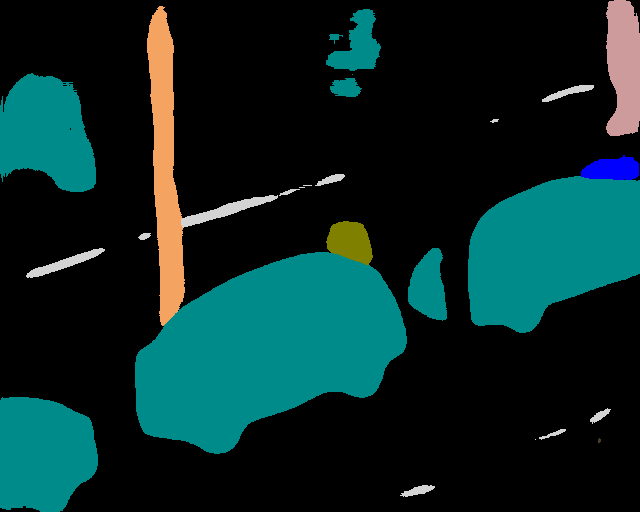} &
        \includegraphics[width=\imgW]{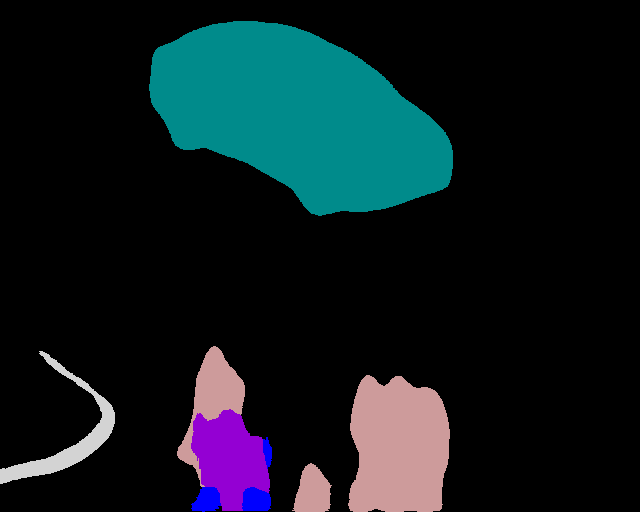} &
        \includegraphics[width=\imgW]{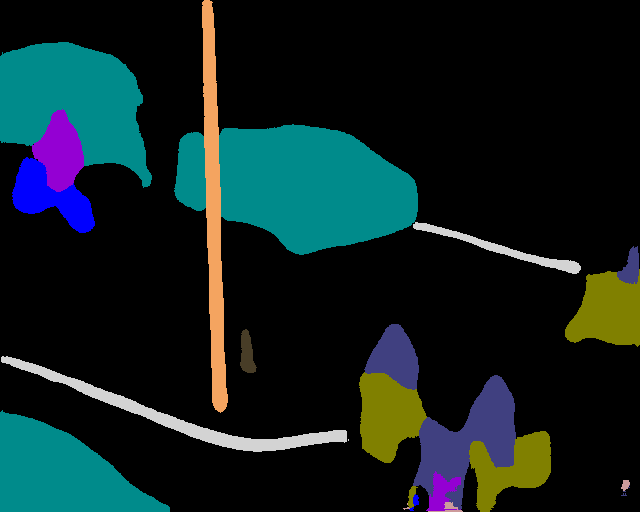} &
        \includegraphics[width=\imgW]{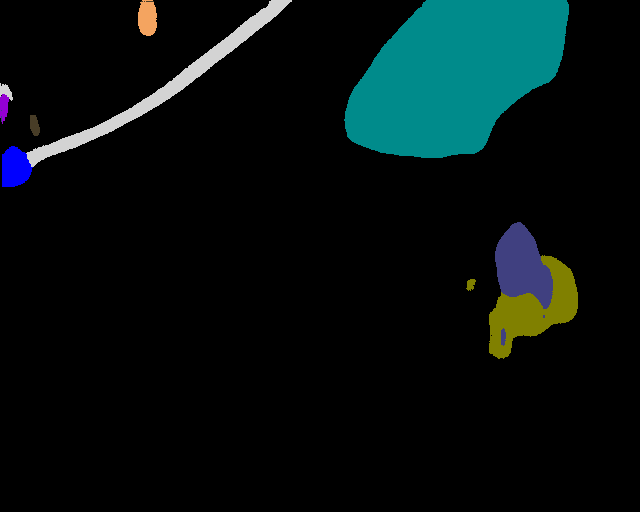} &
        \includegraphics[width=\imgW]{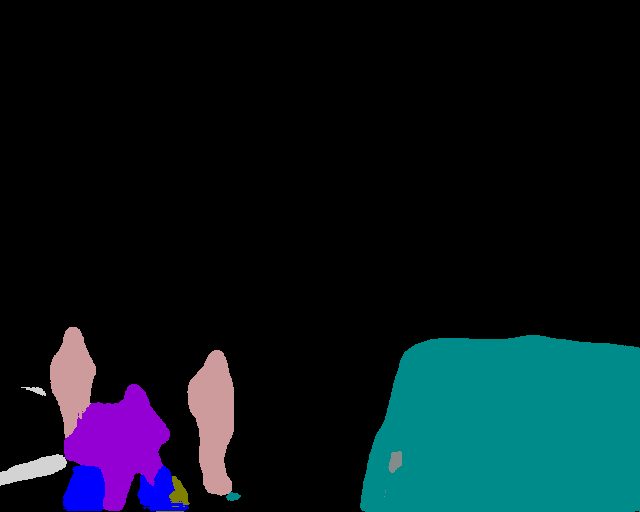} &
        \includegraphics[width=\imgW]{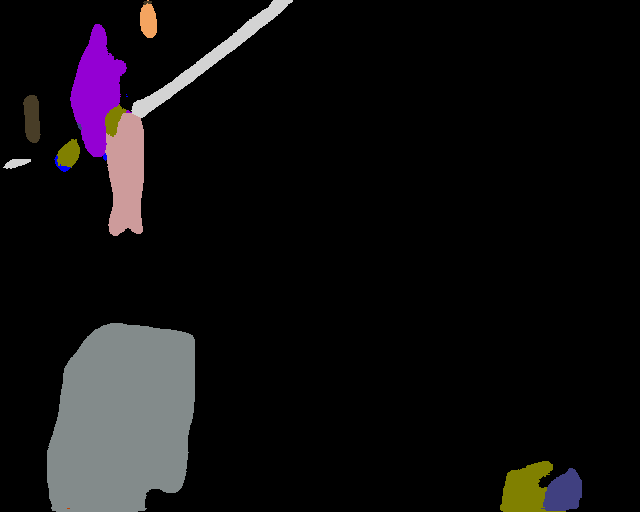} &
        \includegraphics[width=\imgW]{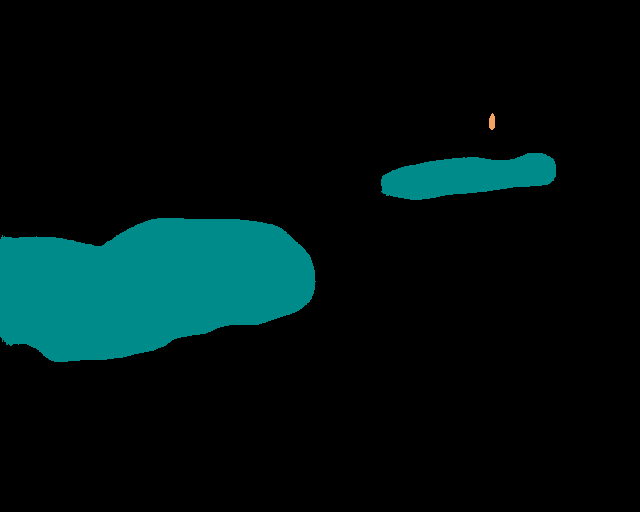} &
        \includegraphics[width=\imgW]{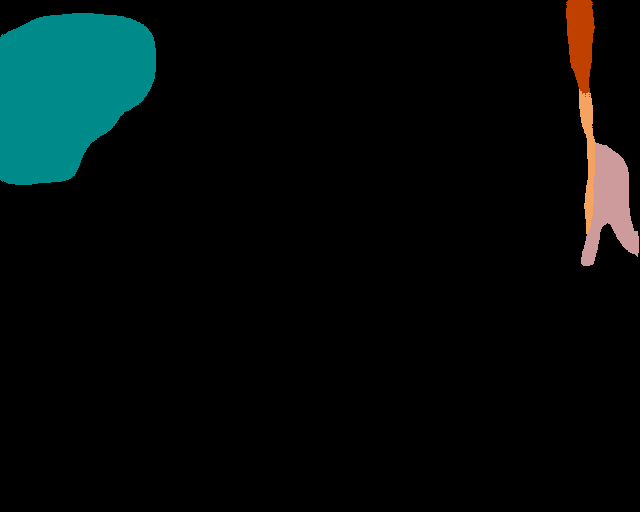} &
        \includegraphics[width=\imgW]{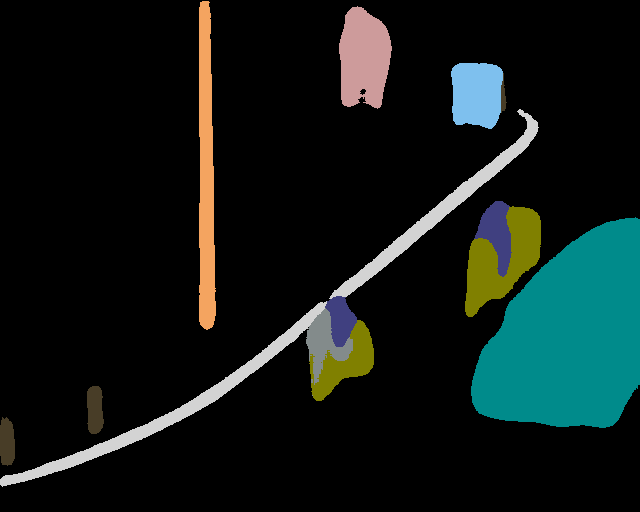} &
        \includegraphics[width=\imgW]{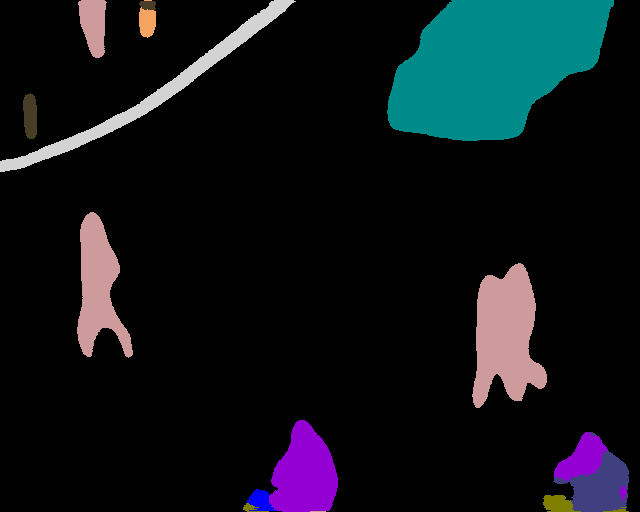} \\

        \rowL{{\fontsize{6}{6.5}\selectfont ViT-\\SemCom-\\OMA}} &
        \includegraphics[width=\imgW]{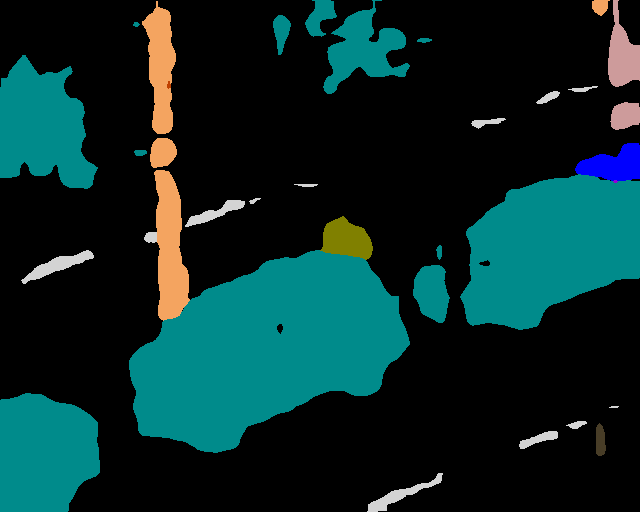} &
        \includegraphics[width=\imgW]{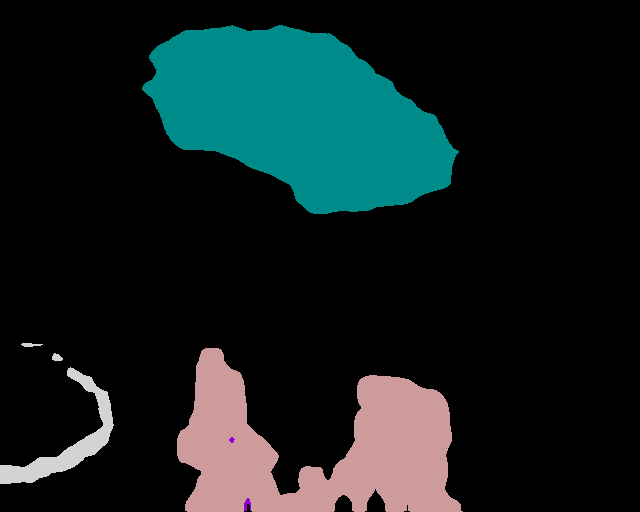} &
        \includegraphics[width=\imgW]{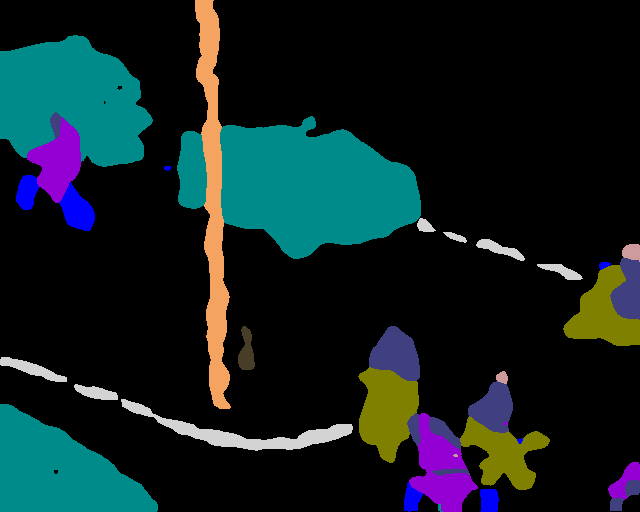} &
        \includegraphics[width=\imgW]{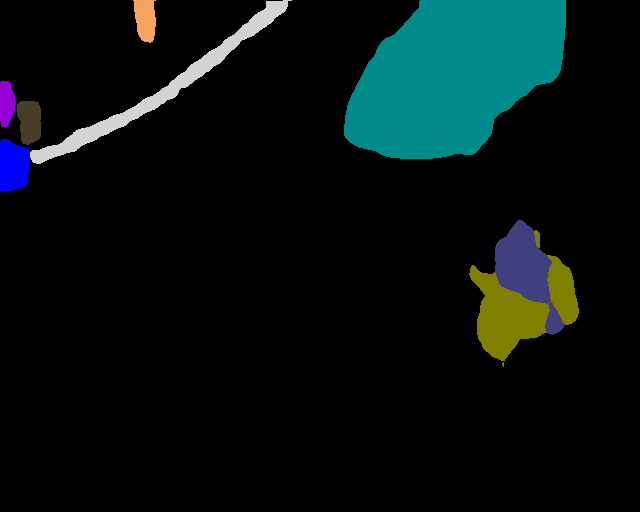} &
        \includegraphics[width=\imgW]{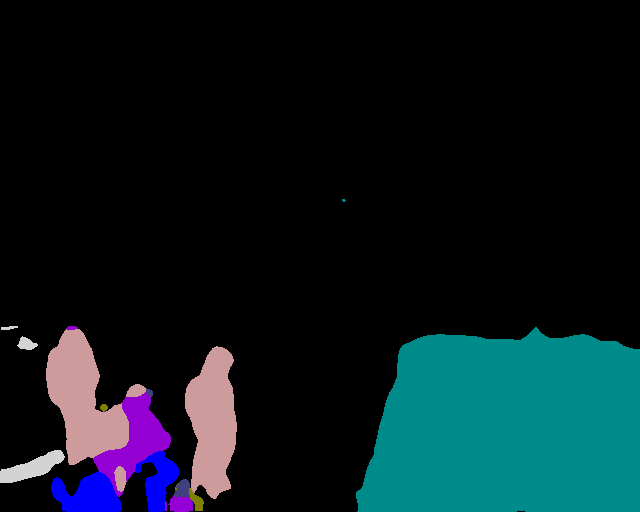} &
        \includegraphics[width=\imgW]{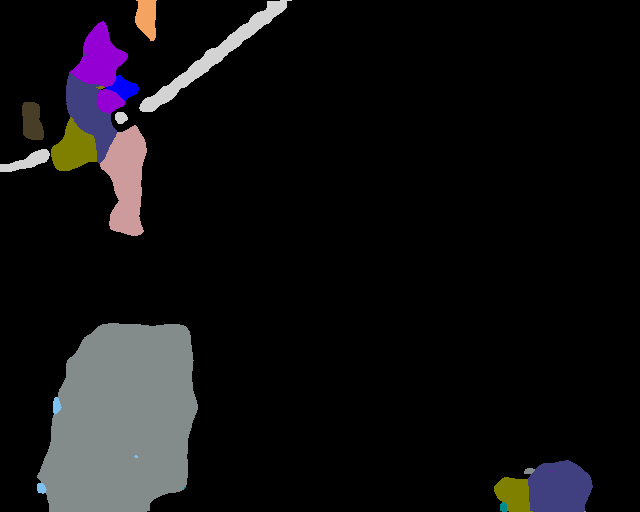} &
        \includegraphics[width=\imgW]{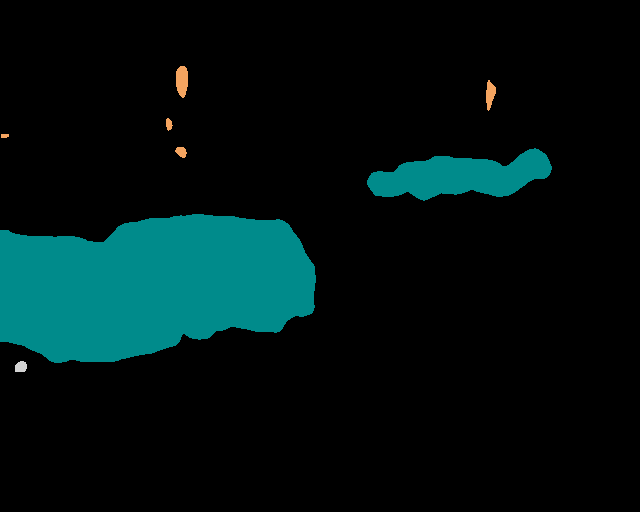} &
        \includegraphics[width=\imgW]{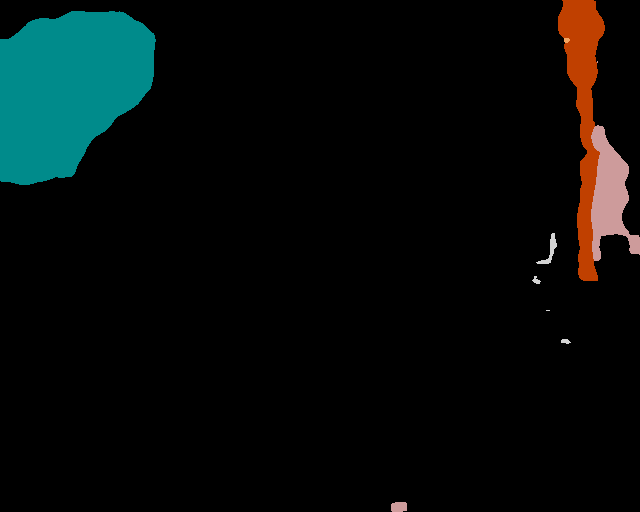} &
        \includegraphics[width=\imgW]{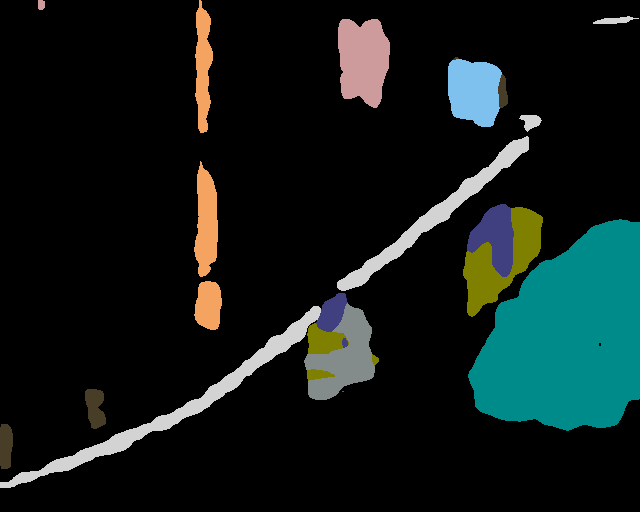} &
        \includegraphics[width=\imgW]{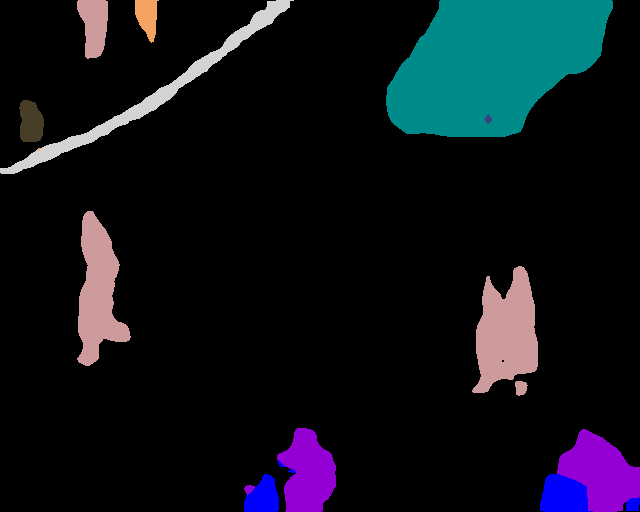} \\
        
    \end{tabular}

    %
    \centering
    \scriptsize 
    \begin{tikzpicture}
        \definecolor{c_carstop}{RGB}{72, 61, 39}
        \definecolor{c_bike}{RGB}{0, 0, 255}
        \definecolor{c_bicyclist}{RGB}{148, 0, 211}
        \definecolor{c_motorcycle}{RGB}{128, 128, 0}
        \definecolor{c_motorcyclist}{RGB}{64, 64, 128}
        \definecolor{c_car}{RGB}{0, 139, 139}
        \definecolor{c_tricycle}{RGB}{131, 139, 139}
        \definecolor{c_trafficlight}{RGB}{192, 64, 0}
        \definecolor{c_box}{RGB}{126, 192, 238}
        \definecolor{c_pole}{RGB}{244, 164, 96}
        \definecolor{c_curve}{RGB}{211, 211, 211}
        \definecolor{c_person}{RGB}{205, 155, 155}

        \fill[c_carstop] (0, 0.3) rectangle (0.3, 0.5); \node[right] at (0.3, 0.4) {CarStop};
        \fill[c_bike] (2.0, 0.3) rectangle (2.3, 0.5); \node[right] at (2.3, 0.4) {Bike};
        \fill[c_bicyclist] (4.0, 0.3) rectangle (4.3, 0.5); \node[right] at (4.3, 0.4) {Bicyclist};
        \fill[c_motorcycle] (6.0, 0.3) rectangle (6.3, 0.5); \node[right] at (6.3, 0.4) {Motorcycle};
        \fill[c_motorcyclist] (8.0, 0.3) rectangle (8.3, 0.5); \node[right] at (8.3, 0.4) {Motorcyclist};
        \fill[c_car] (10.5, 0.3) rectangle (10.8, 0.5); \node[right] at (10.8, 0.4) {Car};
        
        \fill[c_tricycle] (0, 0) rectangle (0.3, 0.2); \node[right] at (0.3, 0.1) {Tricycle};
        \fill[c_trafficlight] (2.0, 0) rectangle (2.3, 0.2); \node[right] at (2.3, 0.1) {TrafficLight};
        \fill[c_box] (4.0, 0) rectangle (4.3, 0.2); \node[right] at (4.3, 0.1) {Box};
        \fill[c_pole] (6.0, 0) rectangle (6.3, 0.2); \node[right] at (6.3, 0.1) {Pole};
        \fill[c_curve] (8.0, 0) rectangle (8.3, 0.2); \node[right] at (8.3, 0.1) {Curve};
        \fill[c_person] (10.5, 0) rectangle (10.8, 0.2); \node[right] at (10.8, 0.1) {Person};

    \end{tikzpicture}
    
    \caption{Visual comparison of semantic segmentation results on the SemanticRT dataset at SNR $=5$ dB and $C=1/512$. We compare the proposed scheme with baseline methods across 10 scenarios. The labels on the left indicate the input modalities (RGB, IR), the ground truth, and predictions from different methods. The color legend at the bottom denotes the semantic classes.}
    \label{fig:visual_segmentation_grid}
    
\end{figure*}

\begin{figure}[!t]
    \centering
    \includegraphics[width=0.96\columnwidth]{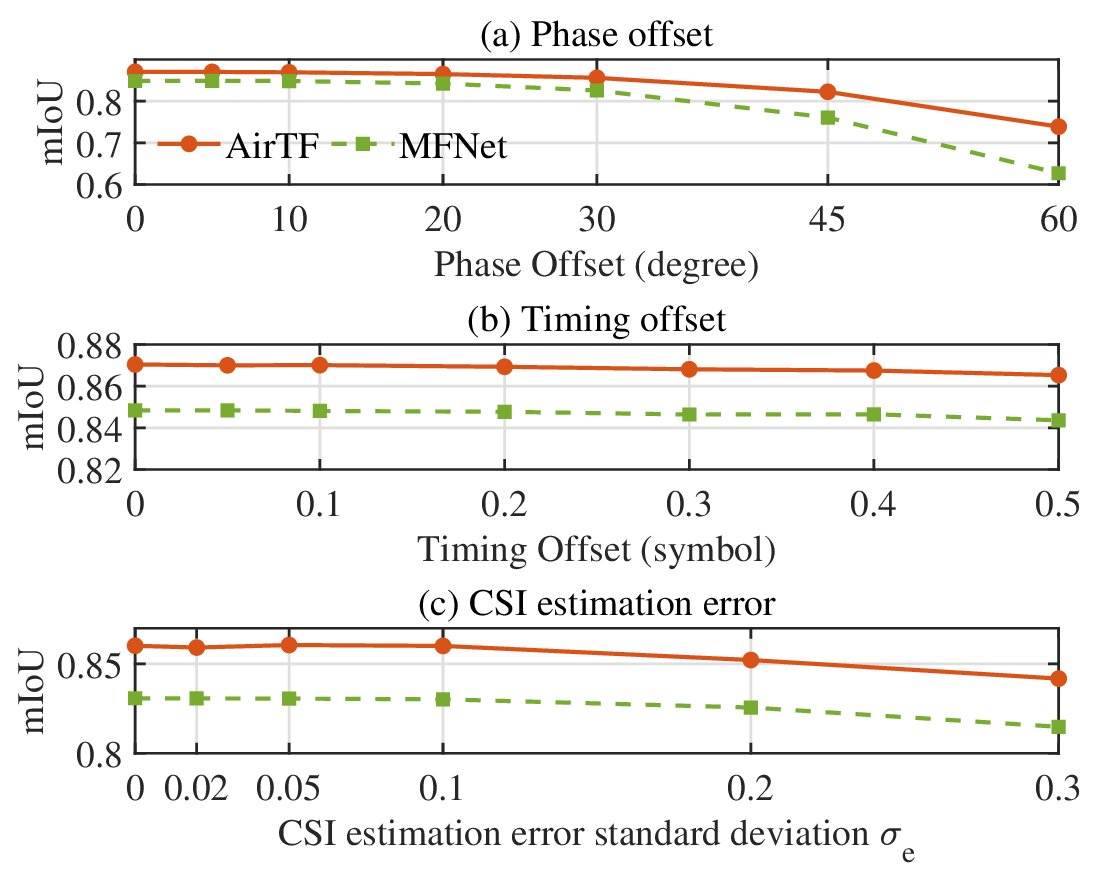}
    
    \caption{Robustness to synchronization imperfection and CSI estimation errors at SNR $=5$ dB and $C=1/256$: (a) residual phase offset; (b) residual timing offset; and (c) CSI estimation error standard deviation.}
    \label{fig:robustness_impairments}
\end{figure}

\subsection{Robustness to Synchronization Imperfection and CSI Estimation Errors}
\label{subsec:robustness}

To examine the sensitivity of AirTF to residual synchronization and CSI estimation errors, we test the trained models under residual phase offsets, residual timing offsets, and imperfect CSI estimation without re-training. As shown in Fig.~\ref{fig:robustness_impairments}(a), AirTF and MFNet both remain stable under moderate phase offsets, while large phase mismatch causes clear degradation. For example, AirTF decreases from 87.04\% mIoU at $0^\circ$ to 85.60\% at $30^\circ$, but drops more noticeably at $60^\circ$.

Fig.~\ref{fig:robustness_impairments}(b) shows that residual sub-symbol timing offsets have limited impact after applying the aligned-sample estimator for misaligned AirComp~\cite{shao2022misaligned,qiao2024massive}. Increasing the residual delay from $0$ to $0.5$ symbol intervals changes the mIoU of both AirTF and MFNet by about 0.5\%, indicating that the tested timing offsets are effectively mitigated. For imperfect CSI estimation, we model the CSI estimate as $\widehat{h}_u=h_u+e_u$, where $e_u\sim\mathcal{CN}(0,\sigma_e^2)$ and $\sigma_e$ denotes the standard deviation of the complex CSI estimation error. Fig.~\ref{fig:robustness_impairments}(c) further shows gradual degradation as $\sigma_e$ increases in Rayleigh fading, while AirTF remains consistently above MFNet across the tested error levels.

We also test moderate spatial calibration errors by perturbing the IR input at SNR $=5$ dB and $C=1/256$. When the IR image is translated by 4 pixels, AirTF obtains 86.23\% mIoU, compared with 84.19\% for MFNet; when the IR image is rotated by $2^\circ$, AirTF obtains 85.34\%, compared with 83.40\% for MFNet. These results indicate that AirTF remains robust to small spatial misalignment, while accurate calibration is still important. {\color{black}Modality-specific encoders accommodate heterogeneous sensing modalities, while deployment across different sensor platforms may additionally require input normalization, sensor calibration, or domain adaptation.}

\subsection{Complexity and Latency}
We further profile the computational cost, communication overhead, and latency at $C=1/256$ with input resolution $512\times640$. Table~\ref{tab:complexity_latency} summarizes the UE-side encoder, ES-side receiver/decoder, and complete forward-pass complexity and latency measured on an NVIDIA RTX 5090 with batch size 16 over 100 repeated runs and reported per image.
\begin{table}[!t]
    \centering
    
    \caption{COMPUTATIONAL COMPLEXITY AND LATENCY COMPARISON.}
    
    \label{tab:complexity_latency}
    \setlength{\tabcolsep}{3pt}

    \footnotesize
    \begin{tabular}{lcccccc}
        \toprule
        \multirow{2}{*}{Model} & \multicolumn{3}{c}{GFLOPs} & \multicolumn{3}{c}{Latency (ms)} \\
        \cmidrule(lr){2-4}\cmidrule(lr){5-7}
        & Enc. & Rx/Dec. & Full & Enc. & Rx/Dec. & Full \\
        \midrule
        AirTF & 171.07 & 585.63 & 756.69 & 8.95 & 7.92 & 16.70 \\
        ViT-SemCom-OMA & 171.06 & 588.67 & 759.72 & 9.01 & 8.12 & 17.14 \\
        MFNet & 11.31 & 8.54 & 19.85 & 0.86 & 0.47 & 1.36 \\
        \bottomrule
    \end{tabular}
    
\end{table}
AirTF and ViT-SemCom-OMA incur higher computation than MFNet due to the ViT encoders and ES-side semantic decoder. {\color{black}AirTF and ViT-SemCom-OMA have comparable complexity and latency for a complete forward pass, indicating that the shared-channel fusion architecture does not introduce additional inference cost relative to the Transformer-based OMA baseline.} Notably, although the AirTF ES-side decoder requires more than three times the FLOPs of its encoder, it incurs a slightly lower latency. This discrepancy reflects the different GPU utilization efficiencies of their underlying operations and indicates that FLOPs alone are not a reliable predictor of wall-clock latency. Therefore, the additional computational cost of AirTF is accompanied by the performance and robustness advantages observed under limited channel bandwidth, fading channels, residual synchronization errors, and imperfect CSI estimation. From the communication perspective, {\color{black}AirTF and ViT-SemCom-OMA use the same total budget of} $N_s=5120$ {\color{black}complex-valued channel uses at $C=1/256$; AirTF uses the shared channel, whereas ViT-SemCom-OMA divides the channel uses between the two modalities. For AirTF, this budget corresponds to} about 0.26 ms over a 20 MHz bandwidth. The two AirTF modality encoders take 8.95 ms in total, i.e., about 4.47 ms per UE when executed as separate branches, indicating that semantic encoding, rather than wireless transmission, dominates the latency.

\subsection{Ablation Study}

To assess the effect of encoder initialization, Fig.~\ref{fig:experimental_results}(d) compares four initialization choices for the RGB and IR ViT encoders. In this ablation, encoders initialized with ImageNet pre-trained weights use a learning rate of $1 \times 10^{-5}$, while randomly initialized encoders use $5 \times 10^{-5}$. The fully random initialization setting is trained for 100 epochs, and the other initialization settings are trained for 50 epochs. At SNR $=5$ dB and $C=1/256$, initializing both encoders from ImageNet achieves 87.05\% mIoU, compared with 82.88\% for random initialization of both encoders. Initializing only the IR encoder from ImageNet reaches 86.40\%, while initializing only the RGB encoder from ImageNet reaches 84.75\%. These results indicate that ImageNet pre-training provides useful visual priors for both modalities, and that jointly adapting both modality-specific encoders gives the most reliable performance.

\subsection{Visual Demonstration}
We first provide qualitative results to visually assess the segmentation quality. Fig.~\ref{fig:visual_segmentation_grid} showcases segmentation outputs under challenging conditions (SNR = 5 dB, $C=1/512$).  AirTF produces segmentation maps that are generally more consistent with the ground truth, preserving more complete object regions and clearer object boundaries. MFNet can recover the main vehicle regions, but it occasionally misses small targets and produces coarser contours. ViT-SemCom-OMA remains visually competitive, while a few local regions show fragmented predictions or loss of fine semantic details. This visual evidence is consistent with the quantitative results, suggesting that AirTF can better preserve object regions and fine semantic details than the baselines under the tested adverse channel conditions.

\begin{figure}[t]
    
    \centering
    \setlength{\tabcolsep}{1pt} 
    \newcommand{\subimgW}{0.30\columnwidth} 
    
    \begin{tabular}{ccc}
        \includegraphics[width=\subimgW]{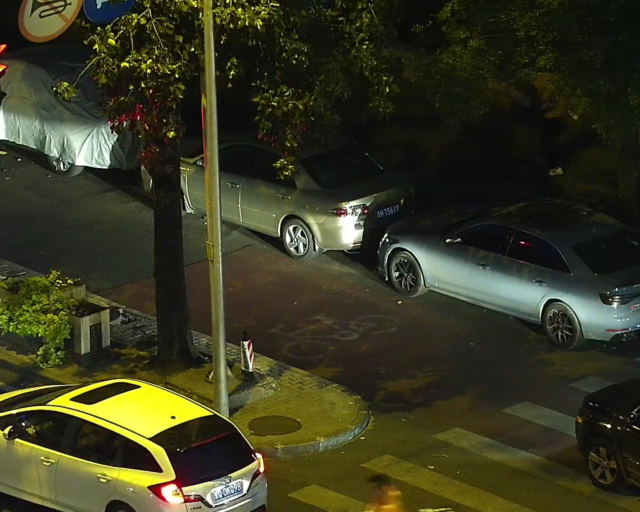} &
        \includegraphics[width=\subimgW]{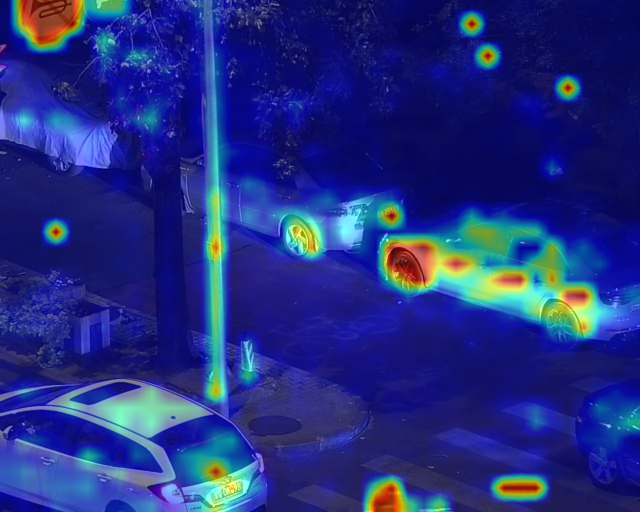} &
        \includegraphics[width=\subimgW]{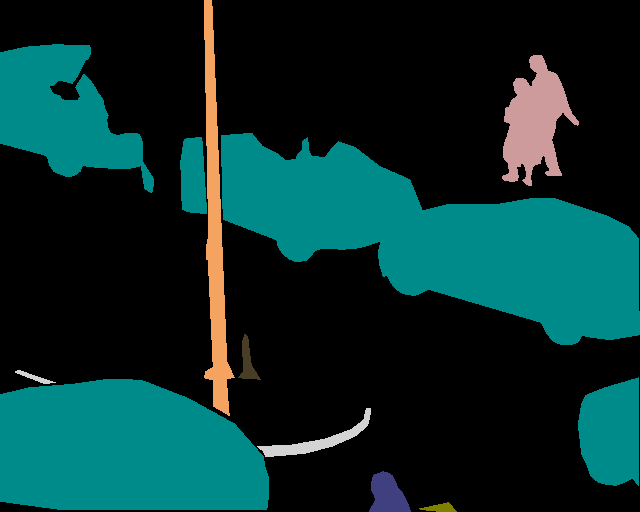} \\
        {\scriptsize (a) Input RGB Image} & {\scriptsize (b) RGB Attention Map} & {\scriptsize (c) Ground Truth} \\

        \includegraphics[width=\subimgW]{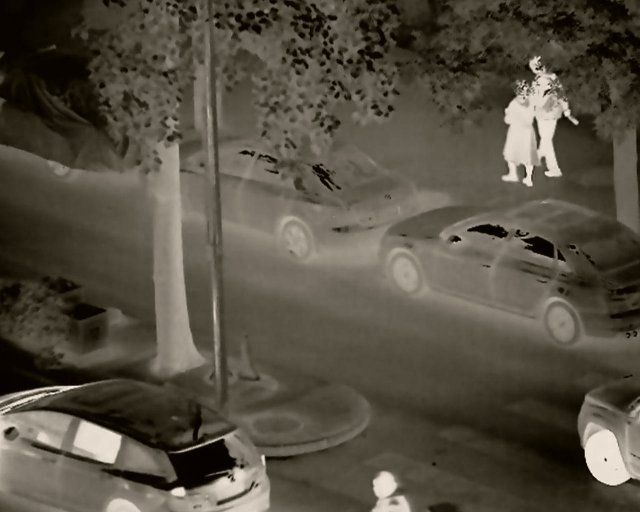} &
        \includegraphics[width=\subimgW]{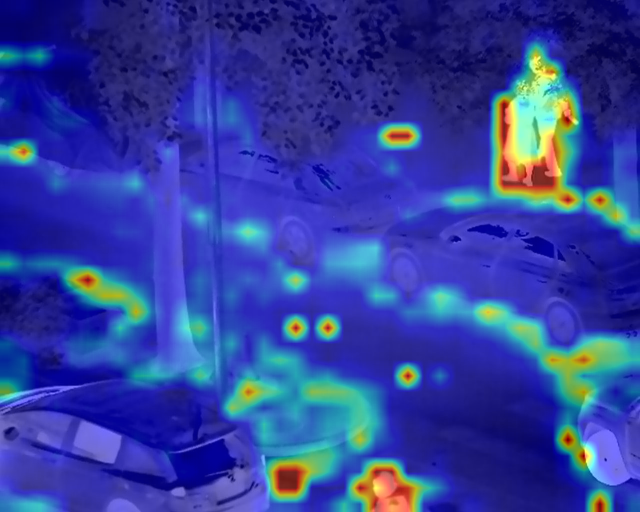} &
        \includegraphics[width=\subimgW]{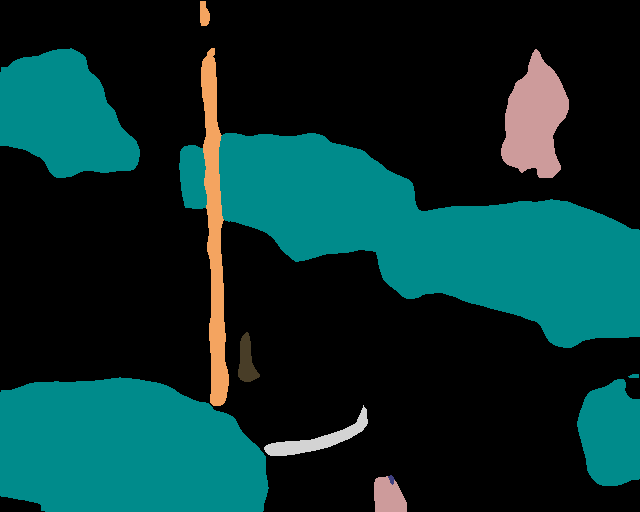} \\
        {\scriptsize (d) Input IR Image} & {\scriptsize (e) IR Attention Map} & {\scriptsize (f) Our Prediction} \\
    \end{tabular}
    
    \caption{Visualization of multi-modal complementarity in the proposed system. High attention scores (red) indicate regions of high semantic importance.}
    \label{fig:attention_vis}
    
\end{figure}

To further interpret how the proposed system effectively fuses multi-modal information via token superposition, we visualize the internal attention mechanisms of the ViT encoders alongside the final segmentation results. Fig.~\ref{fig:attention_vis} presents a qualitative comparison comprising the input modalities, the ground truth, and the model prediction. Specifically, Fig.~\ref{fig:attention_vis}(b) and (e) overlay the self-attention probability maps of the last ViT block onto the input images.

As observed in Fig.~\ref{fig:attention_vis}(b), the RGB branch's attention is primarily focused on visually salient objects with clear textures, such as the pole and the white vehicle. However, the RGB sensor fails to clearly capture some targets in low-light or shadowed regions (e.g., the pedestrians in the top-right corner), rendering them almost invisible in Fig.~\ref{fig:attention_vis}(a). Crucially, the IR branch compensates for this deficiency. As shown in Fig.~\ref{fig:attention_vis}(e), the IR ViT encoder successfully attends to these heat-emitting targets, assigning them high attention scores (red regions) despite the lack of visible illumination. This observation suggests that the two branches provide complementary cues before being superposed over the air. After end-to-end training, the decoder can exploit the received superposed token representation to recover both visually salient structures and thermally visible targets, supporting the interpretation that token-level superposition can serve as a task-oriented semantic fusion mechanism.

\section{Conclusion}
In this paper, we proposed the AirTF framework for task-oriented multi-modal semantic communications. The proposed system utilizes modality-specific ViT encoders to extract semantic tokens from RGB and IR images, which are then concurrently compressed and superposed over a shared wireless channel. With end-to-end task training, the over-the-air superposed token representation can exploit the structural complementarity between modalities, effectively capturing both visually salient textures and thermally visible targets in low-light conditions. Furthermore, initializing the ViT encoders with pre-trained foundation models proves essential, as it provides the necessary visual priors to overcome the data-hungry nature of ViTs on limited-scale semantic segmentation datasets. Experimental results demonstrate that AirTF consistently outperforms the orthogonal token-transmission baseline and the CNN-based fusion baseline across the tested channel conditions, especially under limited channel bandwidth.  Future work will extend AirTF beyond semantic segmentation to broader perception tasks and explore task-adaptive cross-modal token selection to improve the efficiency of multi-sensor semantic transmission.

\IEEEtriggercmd{\balance}
\IEEEtriggeratref{29}
\bibliographystyle{IEEEtran}

\bibliography{references}

\end{document}